\documentclass[twocolumn]{aastex63}

\pdfoutput=1
\usepackage[T1]{fontenc}
\usepackage{graphicx}
\usepackage{amssymb}
\usepackage{gensymb}
\usepackage{amsmath}
\usepackage{bm}
\usepackage{lineno}
\usepackage{chemformula}
\usepackage{booktabs}
\usepackage{multirow}
\usepackage{color}
\usepackage{xcolor}
\usepackage[caption=false]{subfig}

\submitjournal{PSJ}
\begin{document}

\title{Blue marble, stagnant lid: Could dynamic topography avert a waterworld?}

\author{Claire Marie Guimond}
\affiliation{Department of Earth Sciences, University of Cambridge, Downing Street, Cambridge CB2 3EQ, UK}

\author{John F. Rudge}
\affiliation{Department of Earth Sciences, University of Cambridge, Downing Street, Cambridge CB2 3EQ, UK}

\author{Oliver Shorttle}
\affiliation{Department of Earth Sciences, University of Cambridge, Downing Street, Cambridge CB2 3EQ, UK}
\affiliation{Institute of Astronomy, University of Cambridge, Madingley Road, Cambridge CB3 0HA, UK}

\correspondingauthor{Claire Marie Guimond}
\email{cmg76@cam.ac.uk}

\begin{abstract}

Topography on a wet rocky exoplanet could raise land above its sea level. Although land elevation is the product of many complex processes, the large-scale topographic features on any geodynamically-active planet are the expression of the convecting mantle beneath the surface. This so-called ``dynamic topography'' exists regardless of a planet's tectonic regime or volcanism; its amplitude, with a few assumptions, can be estimated via numerical simulations of convection as a function of the mantle Rayleigh number. We develop new scaling relationships for dynamic topography on stagnant lid planets using 2D convection models with temperature-dependent viscosity. These scalings are applied to 1D thermal history models to explore how dynamic topography varies with exoplanetary observables over a wide parameter space. Dynamic topography amplitudes are converted to an ocean basin capacity, the minimum water volume required to flood the entire surface. Basin capacity increases less steeply with planet mass than does the amount of water itself, assuming a water inventory that is a constant planetary mass fraction. We find that dynamically-supported topography alone could be sufficient to maintain subaerial land on Earth-size stagnant lid planets with surface water inventories of up to approximately $10^{-4}$ times their mass, in the most favourable thermal states. By considering only dynamic topography, which has $\sim$1-km amplitudes on Earth, these results represent a lower limit to the true ocean basin capacity. Our work indicates that deterministic geophysical modelling could inform the variability of land propensity on low-mass planets.

\end{abstract}

\section{Introduction}

The concurrence of land and water on a planet's surface will affect its climate state \citep{turbet_habitability_2016, rushby_effect_2019, del_genio_albedos_2019, graham_thermodynamic_2020, zhao_climate_2021}, the planetary context of potential biosignatures \citep{schwieterman_exoplanet_2018, glaser_detectability_2020, lisse_geologically_2020, krissansen-totton_oxygen_2021}, and perhaps its likelihood to host the prebiotic chemistry that leads to the origin of life \citep{patel_common_2015, rimmer_origin_2018, rosas_archaean_2021, van_kranendonk_elements_2021}. Planetary land/ocean fractions emerge in a compromise between water's total budget and distribution between surface and interior reservoirs, and the size of the basins carved out by topography \citep[e.g.,][]{simpson_bayesian_2017}. The resulting ocean mass from the former is largely stochastic: coded within it are the histories of volatile delivery during accretion \citep{raymond_high-resolution_2006, morbidelli_building_2012}, interior degassing from the magma ocean and succeeding mantle \citep{elkins-tanton_linked_2008, schaefer_redox_2017, barth_magma_2020, katyal_effect_2020, ortenzi_mantle_2020, GUIMOND2021106788, lichtenberg_vertically_2021, bower_retention_2021}, atmospheric erosion by impacts \citep{zahnle_cosmic_2017, schlichting_atmosphere_2018, howe_survival_2020}, and photodissociative atmospheric escape \citep{tian_water_2015, zahnle_strange_2019, gronoff_atmospheric_2020}, along with the surface temperature and pressure. In contrast, large-scale aspects of planetary topography may lend themselves to deterministic relationships with observable planetary bulk properties. Although substantial water budgets of a few wt.\% would inevitably produce waterworlds \citep[e.g.,][]{simpson_bayesian_2017}, at smaller water mass fractions the outcome is sensitive to the planet's topography; even a tiny ocean mass would inundate an atopographic body. Early constraints on exoplanet land propensity might therefore start with topography.

This first investigation will limit itself to forms of topography that could exist without moving plates. Whether or not a given planet manifests plate tectonics appears to be hysteretic and largely unanswerable by modelling from state variables \citep{lenardic_notion_2012, weller_evolution_2018, lenardic_diversity_2018}. Consequentially, this paper adopts the working hypothesis that a stagnant lid describes a temperate rocky planet's most natural regime \citep{stern_stagnant_2018}. Here the cool outermost rock layer does not experience enough stress to trigger its breaking into plates by brittle failure.

Of the types of topography on planets, so-called dynamic topography---the surface deformation from convective upwellings and downwellings in the mantle---can create significant elevation differences without requiring plate tectonics. Although dynamic topography is not independent of plate movement on Earth, where mantle convection beneath divergent and convergent plate boundaries has built ridges higher than sea level and trenches deeper than Mount Everest, respectively, and though we expect dynamic topography to be muted in the absence of plate tectonics, mantle convection would retain an inevitable influence on the low-order shape of the stagnant lid surface. That is, dynamic topography is everywhere: a planet exhibits this phenomenon so long as its interior convects. Bodies in our solar system do boast high peaks by other means: massive lava flows (e.g., Olympus Mons) or impact cratering (e.g., Rheasilvia on Vesta). Yet if we are interested in whether a planet's topography could be higher than its sea level \emph{regardless of} volcanism, cratering, and other processes contingent on a planet's specific geological history, then we might begin with dynamic topography as the most endogenously universal of relief mechanisms.

On long length scales of relief, additional support comes from the density contrast between the heavier mantle and lighter crust, which buoys topography at an equilibrium height. This isostatically-compensated topography can be higher in part because the maximum stress underneath the load is shifted to shallower, cooler depths, where the lithosphere is stronger. Parameterisations of isostatic equilibria, however, depend only on the density contrast and thickness of the crust, and so are sensitive to the planet's specific petrologic history. This could be daunting if we consider that the emergence of thick granitic continents on Earth still lacks a consistent explanation, but is probably entwined with its geodynamic history \citep{lenardic_continental_2005,
korenaga_crustal_2018, honing_bifurcation_2019}. Predicting isostatic elevations would require information which may always be model-dependent. Purely dynamic topography, meanwhile, both originates from and is supported by the sole process of thermal convection. It is directly obtained from any numerical convection model \citep[e.g.,][]{mckenzie_surface_1977, kiefer_geoid_1992, kiefer_geoid_1998, huang_constraints_2013, arnould_scales_2018, lees_gravity_2020}; its prediction requires less prior knowledge.

\begin{figure*}
\gridline{\fig{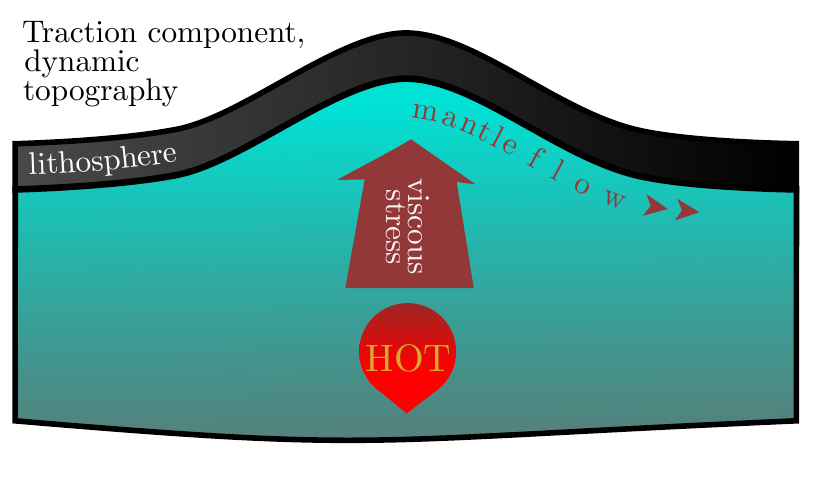}{0.5\textwidth}{(a)}
          \fig{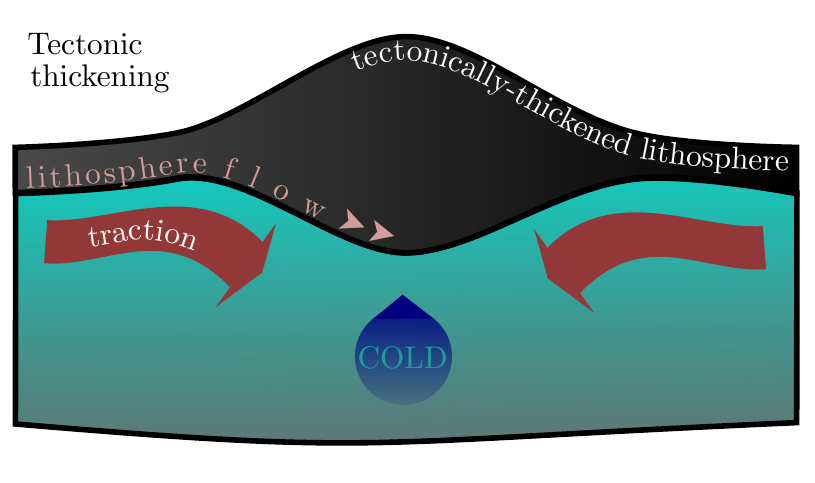}{0.5\textwidth}{(b)}}
          \gridline{\fig{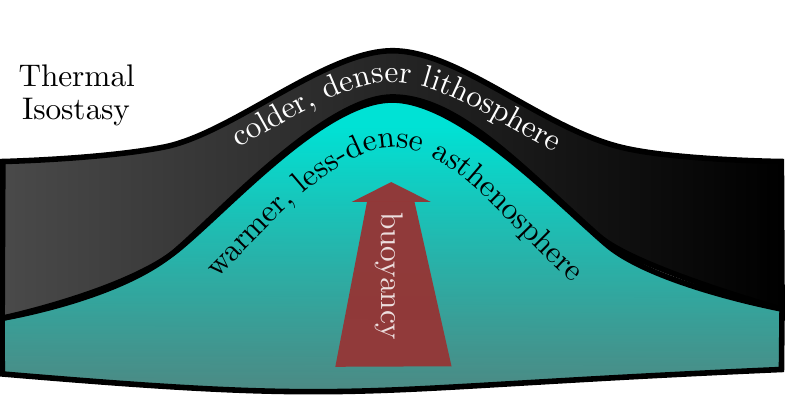}{0.5\textwidth}{(c)}
          \fig{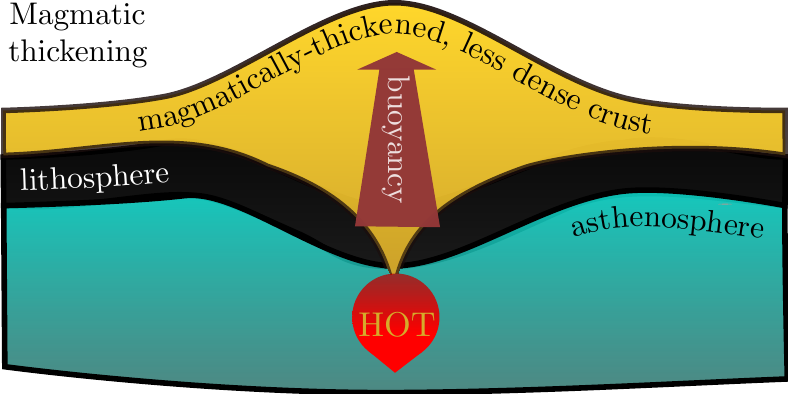}{0.5\textwidth}{(d)}}
          
\caption{The four major endogenic sources of topography on a stagnant lid planet. \textit{(a)} The component of dynamic topography due to flow-induced traction on the lithosphere. \textit{(b)} Tectonic crustal thickening caused by tension over cold downwellings. \textit{(c)} The component of dynamic topography due to thermal isostasy over thinned lithosphere. \textit{(d)} Magmatic crustal thickening caused by melting of upwelling plumes.}
\label{fig:topography-schematic}
\end{figure*}

Note that stagnant lid convection can lead to other forms of topography, beyond just that supported dynamically by convection (figure \ref{fig:topography-schematic}). The melting associated with hot upwelling mantle can form thick, low-density crust as in figure \ref{fig:topography-schematic}d \citep{stofan_large_1995}; further, tension above downwelling plumes can also thicken the crust tectonically as in figure \ref{fig:topography-schematic}b \citep{kiefer_mantle_1991,pysklywec_time-dependent_2003, zampa_evidence_2018}. Both phenomena would induce compositional isostasy, resulting in altitudes unrepresentative of pure dynamic support. Neither, however, will be included in the groundwork we perform here. There is also a distinction to be made for \textit{thermal} isostasy, in which thermal expansion of the lithospheric mantle creates the density difference, rather than compositional separation related to melting (figure \ref{fig:topography-schematic}c). Hot upwelling mantle will induce thermal isostasy. By convention, we do include thermal isostasy within the full dynamic topography \citep[see][]{molnar_mantle_2015, hoggard_observational_2021}. Overall, the elevations we model here should represent conservative lower limits on a stagnant lid planet's static topography.

In summary, among the large-scale mechanisms sculpting the surface of an active planet, dynamic topography alone has the advantage of being (i) inevitably present, regardless of tectonic mode; and (ii) a direct result, quantitatively, of a tractable process (mantle convection). From a modelling perspective, all of these factors help define a simplified and tractable problem: how does dynamic topography scale with parameters that dictate how a planet will convect---like the depth of the mantle, the thermal state, or the rheology? In principle this scaling relationship can be extracted from numerical simulations of convection. From there, cheaper 1D parameterised convection models can use this scaling to explore how the amplitude of dynamic topography changes over a wide range of planetary bulk properties. Because the scaling itself may be sensitive to a planet's tectonic mode, our convection simulations neglect the possibility of plates.

\subsection{Dynamic topography scaling relationships}\label{sec:intro-scaling}

Limited by computing power, early constructions of a scaling function for dynamic topography have used a constant viscosity for the convecting region \citep{parsons_relationship_1983, kiefer_geoid_1992}. Under this isoviscous paradigm, a single dimensionless parameter, the Rayleigh number, describes the convective vigour of the system:
\begin{equation}\label{eq:Ra}
    {\rm Ra} = \frac{\alpha\rho g \Delta T d^3}{\kappa \eta},
\end{equation}
where $\alpha$ is the thermal expansivity of the material in K$^{-1}$, $\rho$ is its density in ${\rm kg\,m^{-3}}$, $g$ is the surface gravity in ${\rm m\,s^{-2}}$, $\Delta T$ is the temperature contrast across the layer in K, $d$ is the depth of the convecting region in m, $\kappa$ is the thermal diffusivity in ${\rm m^2\,s^{-1}}$, and $\eta$ is the dynamic viscosity in Pa~s---in isoviscous convection these parameters are all constant. The Ra number can act as a useful independent scaling variable for many convection phenomena because the vast majority of temperature variations in a convecting cell occur in its boundary layers \citep{mckenzie_convection_1974}. Boundary layer theory justifies a power-law relationship between Ra and the thickness of the upper thermal boundary layer. Hence these previous works on isoviscous dynamic topography supposed scaling relationships of the form $h/(\alpha \Delta T d) \sim {\rm Ra}^n$ (the $\alpha \Delta T d$ term ensures that both sides of the proportionality are dimensionless and $n$ is uniquely defined). 

In rocky planets, however, $\eta$ changes with temperature \citep{karato_rheology_1993}; steep viscosity gradients across the mantle are a defining trait of natural stagnant lid convection in that the cold surface is too viscous to flow \citep{davaille_transient_1993, solomatov_scaling_1995}. Scalings based on (\ref{eq:Ra}) defined using constant viscosity will not necessarily provide an optimal fit to the topography of stagnant lid bodies \citep{sembroni_impact_2017, bodur_impact_2019}. In identifying a convecting system whose viscosity decreases quickly with temperature, we need a second dimensionless parameter in addition to a Rayleigh number: the viscosity contrast across the layer, $\Delta \eta=\eta_0/\eta_1$, where $\eta_0$ is the viscosity at the top and $\eta_1$ the viscosity at the bottom. A nonuniform viscosity profile implies many possible thermal Rayleigh numbers. Here Ra$_1$ denotes (\ref{eq:Ra}) evaluated at $\eta=\eta_1$. In simple numerical models, viscosity is often assumed to follow an exponential law, $\eta(T) = \eta_0 \, {\rm exp}(-b T)$, where the temperature prefactor $b = \ln\left(\Delta \eta\right)$ is a constant \citep{solomatov_scaling_1995}. 

Further, any Ra scaling function will only apply to its intended convection regime. Canonical studies of temperature-dependent viscosity convection distinguish between at least two series of regimes. These regimes have their own transitions in Ra$_1$-$\Delta \eta$ space, which would manifest as discontinuities in the scaling function. 
A first series concerns the mobility of the surface: as $\Delta \eta$ increases, a convecting system will move from a small viscosity contrast regime (similar to the isoviscous case) to a stagnant lid regime, via an intermediate regime of a sluggish lid \citep{solomatov_scaling_1995, moresi_numerical_1995, kameyama_transitions_2000}. In a second series of transitions, the so-called stationarity of convection changes. As Ra$_1$ increases, the system will move roughly from a steady-state regime to a chaotic time-dependent regime, again through a transitional regime \citep{dumoulin_heat_1999, solomatov_scaling_2000}. For either series, the regime boundaries are not sharp in Ra$_1$-$\Delta \eta$ space, but depend on this parameter space in a complex way via the aspect ratio of convection and the initial conditions. Whilst ascribing Ra$_1$ presumes a bottom-heated convection cell, different modes of heating may also affect dynamic topography scaling relationships in ways we do not yet understand. 

A waypoint objective of this work is therefore to develop a preliminary dynamic topography scaling relationship for the stagnant lid regime. Whilst the topography of  stagnant lid bodies has indeed been modelled numerically before \citep{moresi_interpreting_1995, solomatov_stagnant_1996, vezolainen_timing_2003, vezolainen_uplift_2004, orth_isostatic_2011, golle_topography_2012, huang_constraints_2013}, the majority of this literature is directed at producing geoid-to-topography ratios to invert for interior properties of Venus or Mars, as opposed to fully exploring parameter space with forward models. As such, we are aware of no published scalings as explicit functions of the relevant convective parameters. Given the scope of our work here, we do not attempt to characterise the scaling behaviour near the regime discontinuities (which would require a much finer grid of models in Ra$_1$-$\Delta \eta$ space). Instead, we restrict ourselves to the chaotic time-dependent stagnant lid regime located in \citet{moresi_numerical_1995} and \citet{orth_isostatic_2011}, and simulated previously with Venus- and Mars-like parameters \citep{solomatov_scaling_2000, hauck_thermal_2002, reese_scaling_2005, orth_isostatic_2011}. As such, we are assuming that chaotic stagnant lid convection will apply to most geodynamically-active rocky planets---an assumption that may be tested in the future when detailed characterisation of rocky exoplanets becomes possible.

\subsection{The harmonic structure of planetary surface relief} \label{sec:intro-spectrum}

In the second part of this study, we convert scaled dynamic topographies into the corresponding volumes of the largest possible ocean basin. The key product here is a spherical harmonic expansion of this scaled topography onto a Cartesian grid, as a synthetic elevation map. Yet, all that our stagnant lid convection scaling law provides is a scalar height value. With some convenient assumptions about dynamic topography's spectral properties, it is in fact straightforward to find a power spectrum which is consistent with both the scalar height we have, and with some set of spherical harmonic coefficients we need.

Initial observations of Venus, Earth, and Mars' (total) topographies suggested a remarkably log-linear relationship between the 1D power spectral density, $\phi^{\rm PSD}_h$ in m$^3$, and the wavenumber, $k$ in m$^{-1}$. From spherical harmonic degree $l=5$ to at least $l=100$, the available spectra appeared consistent with a slope dlog$\phi^{\rm PSD}_h$/dlog$k \sim-$2 \citep{turcotte_fractal_1987, rapp_decay_1989, balmino_spectra_1993}. This precise slope value was predicted earlier still by \citet{vening1951remarkable} and appears to be physically-motivated \citep{sayles_surface_1978, lovejoy_l12_1995}---perhaps emerging from sediment transport laws \citep{pelletier_why_1997, pelletier_self-organization_1999, roberts_generation_2019}, although we will not be considering topography's modification by erosion explicitly. Statistically, a slope of $-$2 corresponds to red noise, the noise associated with a random walk process.

The convenient consequence of a log-linear spectral model---with a pre-determined slope---is that it would let us approximate the shape of any planetary surface given just one free parameter; i.e., the $y$-intercept of $\phi^{\rm PSD}_h(k)$. 
As for dynamic topography in particular, models and Earth observations have indicated a shallower spectral slope roughly consistent with pink noise $\propto k^{-1}$, up to $l\sim30$ \citep{hoggard_global_2016, hoggard_oceanic_2017, davies_earths_2019}. However, there is no evidence that this spectral structure should characterise dynamic topography under all tectonic regimes. Hence, we extract the spectral structure of our own numerically-modelled stagnant lid topography profiles. We will see that our rudimentary analysis again produces constant dlog$\phi^{\rm PSD}_h$/dlog$k$ values, albeit ones more strongly negative than $-$2. Observations of real stagnant lid bodies in the solar system could then suggest an empirical modification of this purely-dynamic spectral model.

\subsection{Study outline}

Our methods are described in section \ref{sec:methods}. The approach we take is outlined as follows: we begin by extracting scaling relationships for the RMS amplitude of dynamic topography from 2D numerical mantle convection simulations with temperature-dependent viscosity (section \ref{sec:results-scaling}). Second, we embed these scaling relationships in a suite of 1D parameterised convection models, allowing us to explore the sense of change of RMS dynamic topography across a wide model parameter space and planet age distribution (section \ref{sec:results-parameters}). For this parameter study we focus on the planet mass, age, radiogenic element abundance, and core mass fraction, all relevant to the cooling history and Rayleigh number of a planet. We focus on these four parameters because they may be amenable to being observationally constrained for exoplanets, at least in principle. Third, we synthesise 2D maps from the projected RMS amplitudes to see how the maximum capacity of ocean basins, and hence the minimum elevation gain needed for dry land, might trade off with planet size (section \ref{sec:results-ocean}). We end with a discussion of the study's limitations and applicability (section \ref{sec:discussion}).

\section{Methods} \label{sec:methods}

\subsection{Numerical convection model} \label{sec:methods-numerical}

Numerical computations were performed using the ASPECT code version 2.2.0 \citep{KHB12,heister_high_2017, bangerth_aspect_2020}. For each case we systematically varied two key input parameters: Ra$_1$ and $\Delta \eta$. Although we originally explored Ra$_1$ varying from $1 \times 10^7$ to $3 \times 10^8$, we found that simulations below Ra$_1 = 1 \times 10^8$ were not in the chaotic time-dependent regime, and showed characteristically different topography scaling behaviour. Because the present study was not designed to precisely locate these transitions, we focused only on the chaotic time-dependent regime. Simulations above Ra$_1 = 3 \times 10^8$ were found to be computationally impractical.

Our ASPECT implementation 
results in dimensionless temperature and velocity fields, denoted by the prime symbol. These and their derivative quantities can be dimensionalised as, e.g.,
\begin{align}
    \begin{split} \label{eq:dimensionalise}
    T &= \Delta T \; T^\prime + T_0, \\
    u &= \frac{\kappa}{d} \; u^\prime, \\
    (x, y) &= d \; (x^\prime, y^\prime), \\
    \delta_{\rm rh} &= d \; \delta_{\rm rh}^\prime, \\
    \delta_{\rm lid} &= d \; \delta_{\rm lid}^\prime, \\
    h &= \alpha \Delta T d \; h^\prime,
    \end{split}
\end{align}
where $T^\prime$ is the dimensionless temperature, $u^\prime$ is the horizontal component of the dimensionless velocity, $\delta_{\rm rh}^\prime$ is the dimensionless thickness of the upper thermal boundary layer, $\delta_{\rm lid}^\prime$ is the dimensionless thickness of the stagnant lid, $h^\prime$ is the dimensionless height of topography, $T_0$ is the dimensional surface temperature, $\Delta T$ is the dimensional temperature difference from bottom to surface, and the other (dimensional) parameters are defined under (\ref{eq:Ra}) above. 

All simulations involve a 2D rectangular box with fixed top and bottom temperatures, $T_0^\prime = 0$ and $T_1^\prime = 1$ respectively, and no internal heating. Free-slip boundary conditions are ascribed to the top and bottom surfaces, whilst reflecting boundary conditions are ascribed to the sides. We use a wide box with a nondimensional depth $Y^\prime$ of unity and a nondimensional width of $X^\prime = 8Y^\prime$ to minimise the influence of the side walls. We assume an incompressible, infinite-Prandtl-number fluid and use the Boussinesq approximation. Viscosity is Newtonian and varies with temperature according to an exponential rheology law, $\eta^\prime = \exp(-b \, T^\prime)$, where $b = \ln (\Delta \eta)$. We use the coarsest mesh size still able to resolve the lower thermal boundary layer; this varies for different Ra$_1$. Table \ref{tab:aspect} lists the relevant details of the model setup.

Each experiment is deemed to have reached quasi-steady-state when both its RMS velocity stabilises to within 0.002\% and its top and bottom heat fluxes converge. All time steps prior to this point are discarded, and the models are then allowed to run for long enough such that the distribution of RMS dynamic topography is well-characterised. All cases are confirmed to be in the stagnant lid mode of convection based on the surface mobility criterion, $S = (\delta_0^\prime)^2 u_0^\prime \ll 1$, where $\delta_0^\prime = \delta_{\rm lid}^\prime + \delta_{\rm rh}^\prime$ is the dimensionless thickness of the lithosphere, and $u_0^\prime$ is the dimensionless surface velocity \citep{solomatov_three_1997}.

\begin{table}
\centering
\caption{Numerical model setup.\label{tab:aspect}}
\footnotesize
\begin{tabular}{@{} c r r r l @{}}
\toprule
Case & Ra$_1$ & $\Delta \eta$ & Mesh size & Initial temperatures \\
\midrule

1 & $1 \times 10^8$ & $1 \times 10^6$ & 512 $\times$ 64 & Sinusoid \\
2 & $2 \times 10^8$ & $1 \times 10^6$ & 1024 $\times$ 128 & Sinusoid \\
3 & $3 \times 10^8$ & $1 \times 10^6$ & 1024 $\times$ 128 & Sinusoid \\

4 & $1 \times 10^8$ & $1 \times 10^7$ & 512 $\times$ 64 & Sinusoid \\
5 & $2 \times 10^8$ & $1 \times 10^7$ & 1024 $\times$ 128 & Case 4 \\
6 & $3 \times 10^8$ & $1 \times 10^7$ & 1024 $\times$ 128 & Case 4 \\

7 & $1 \times 10^8$ & $1 \times 10^8$ & 1024 $\times$ 128 & Case 4 \\
8 & $2 \times 10^8$ & $1 \times 10^8$ & 1024 $\times$ 128 & Case 4 \\
9 & $3 \times 10^8$ & $1 \times 10^8$ & 1024 $\times$ 128 & Case 4 \\

10 & $1 \times 10^8$ & $1 \times 10^9$ & 1024 $\times$ 128 & Case 4 \\
11 & $2 \times 10^8$ & $1 \times 10^9$ & 1024 $\times$ 128 & Case 4 \\
12 & $3 \times 10^8$ & $1 \times 10^9$ & 1024 $\times$ 128 & Case 4 \\

\bottomrule
\end{tabular}
\end{table}

\subsubsection{Extraction of parameters from the temperature and velocity profiles}  \label{sec:T_retrieval}

The average thickness of the stagnant lid, $\delta_{\rm lid}^\prime$, is found using the graphical method of \citet{solomatov_scaling_2000}. We first fit a smoothing spline of degree $4$ to the horizontally-averaged, time-averaged velocity magnitude profile. To ensure we are detecting the lid, we find the inflection point associated with the greatest velocity magnitude, and ignore the region downwards of this point. We then find the maximum gradient of the remaining spline. The intersection of the depth ($y^\prime$) axis with the tangent to the maximum gradient locates the base of the lid, $y^\prime_{\rm lid}$, so $\delta_{\rm lid}^\prime = Y^\prime - y^\prime_{\rm lid}$. 

Another degree-4 spline fit to the temperature profile, also horizontally-averaged and then time-averaged, tells us the lid basal temperature $T_{\rm lid}^\prime$, being the value of the spline at $y^\prime_{\rm lid}$. The temperature of the nearly-isothermal interior, $T_i^\prime$, is defined by \citet{solomatov_scaling_2000} as the local maximum horizontally-averaged temperature in the convecting layer. Here, we systematically interpret this local maximum as the uppermost inflection point in the temperature spline. 

Immediately below the stagnant lid is the upper thermal boundary layer. Unlike the cold lid, this thinner layer is dynamically unstable and does interact with the rest of the convection cell; cold downwellings form locally where its thickness exceeds a critical value. 
Its thickness is given by $\delta_{\rm rh}^\prime = (T_i^\prime - T_{\rm lid}^\prime)/F_0^\prime$, where $F_0^\prime$ is the total dimensionless heat flux out of the upper boundary divided by $X^\prime$ \citep{thiriet_scaling_2019}. The drop from $T_i^\prime$ to $T_{\rm lid}^\prime$ defines $\Delta T_{\rm rh}^\prime$, the temperature contrast across the upper thermal boundary layer. The commonplace subscript denotes ``rheological" because $\Delta T_{\rm rh}^\prime$ is tied to the rate of change of $\ln(\eta)$ with temperature; in exponential viscosity models this is always a constant and proportional to $b$.

\subsubsection{Fitting a topography scaling relationship} \label{sec:methods-hscaling}

The ASPECT code calculates the horizontal profile of the surface dynamic topography via a stress balance at the centre of each cell on the top boundary,
\begin{equation} \label{eq:h_aspect}
    \sigma_{yy} = - \rho g h,
\end{equation} 
where $\sigma_{yy}$ is the vertical component of the stress imparted by convection, $g$ is the gravity, and $\rho$ is the mantle density. Equation (\ref{eq:h_aspect}) assumes mechanical equilibrium between the surface topography and the interior density structure, a safe assumption for the long timescales of convection \citep[e.g.,][]{ricard_physics_2015}. At each time step, we first normalise the dimensionless topography profile to ensure its mean is zero, and then find its RMS value, $h_{\rm rms}^\prime$. 

We choose the RMS amplitude of topography as the representative scalar quantity to fit, rather than the peak amplitude. This choice is based on the reasoning that the RMS value may be less sensitive to the model geometry---crucial for inferring 3D behaviour from 2D models, as we will be doing. As such, we ran preliminary isoviscous convection simulations to confirm that neither Cartesian nor cylindrical 2D geometries show the same peak topographies as the equivalent 3D spherical experiments from \citet{lees_gravity_2020}, whereas, for all three setups, the RMS topographies align well. That the same result holds for non-isoviscous simulations is an outstanding caveat of this study.

Earlier in section \ref{sec:intro-scaling}, we motivated the need for two parameters, a Rayleigh number and viscosity contrast, to fully describe stagnant lid convection. These will serve as the independent variables in the scaling function. 
We define an interior Rayleigh number,
\begin{equation}\label{eq:Ra_i}
    {\rm Ra}_i = \frac{\alpha\rho g \Delta T d^3}{\kappa \eta(T_i)} = {\rm Ra_1} \frac{\eta(T_1)}{\eta(T_i)};
\end{equation}
that is, evaluating (\ref{eq:Ra}) using the ``interior'' viscosity at $T_i^\prime$  \citep{solomatov_scaling_2000}. This formulation of the Rayleigh number is easily transferable to 1D convection models that predict a single mantle temperature, and sidesteps any problems with predicting lower mantle viscosities (where pressure effects are important). Also with an eye toward 1D model integration, we use the exponential temperature prefactor $b = \ln (\Delta \eta)$ as the second variable. We anticipate a power-law relationship and thus fit a linear model to $b$, $\log({\rm Ra}_i$), and $\log(h_{\rm rms}^\prime$), with an interaction term between $b$ and $\log({\rm Ra}_i$):

\begin{equation} \label{eq:h_Ra_scaling}
\log h_{\rm rms}^\prime = A  + B b + C\log{\rm Ra}_i + D \left(b  \log{\rm Ra}_i\right),
\end{equation}
where $T_i^\prime$ in (\ref{eq:Ra_i}) is determined from the horizontally- and time-averaged temperature profile as per section \ref{sec:T_retrieval}, $h_{\rm rms}^\prime$ is taken as the mean of the RMS value over all time steps, and the log notation refers to the base-10 logarithm here and throughout. Thus, each experiment provides one ($b$, Ra$_i$, $h_{\rm rms}^\prime$) coordinate. Whilst these data have some distribution due to the chaotic time-dependence of convection, we found that including the standard error of the mean of $\log h_{\rm rms}^\prime$ has negligible effect on the regression results (for simplicity we do not consider the uncertainty on Ra$_i$). 

Coefficients $A$, $B$, $C$, $D$, and their covariance matrix are estimated using orthogonal distance regression. The interaction term, $D \left(b  \log{\rm Ra}_i\right)$, accounts for cross-effects between $b$ and Ra$_i$. Although including the interaction term adds an extra parameter, we will see that we need this term to properly capture the observed effect of Ra$_i$ on $h_{\rm rms}^\prime$, which has magnitude and direction depending strongly on $b$ as the data will show; the presence of the fourth term decreases the residual variance of the fit by three-fold compared to its absence.

\subsection{Parameterised thermal history model}

\begin{table*}
\centering
\caption{Dimensional parameters used in the 1D thermal history model. The top panel lists parameters which are constant in all runs. The middle panel lists those parameters which are systematically varied in certain sections of the study, and held constant at the baseline value where noted. The bottom panel lists the unknowns, treated here as random variables distributed as given, such that a distribution of output parameters is obtained.  \label{tab:params}}

\footnotesize
\begin{tabular}{@{} c l r l p{6cm} @{}}
\toprule
Symbol & Description & Value & Units & Ref. \\
\midrule
\multicolumn{5}{c}{\textbf{Constant bulk properties for all planets}} \\
$\rho_m$ & Mantle density & 3500 & $\rm{kg\,m^{-3}}$ &  \citet{thiriet_scaling_2019}  \\
$c_m$ & Mantle specific heat & 1142 & $\rm{J\,kg^{-1}\,K^{-1}}$ & \citet{thiriet_scaling_2019}  \\
$c_c$ & Core specific heat & 840 & $\rm{J\,kg^{-1}\,K^{-1}}$  & \citet{thiriet_scaling_2019}  \\
$k_m$ & Mantle thermal conductivity & 4 & $\rm{W\,m^{-1}\,K^{-1}}$ & \citet{thiriet_scaling_2019}  \\
$\alpha_m$ & Mantle thermal expansivity &  $2.5 \times 10^{-5}$ & K$^{-1}$  & \citet{thiriet_scaling_2019}  \\
$\kappa_m$ & Mantle thermal diffusivity &  $1 \times 10^{-6}$ & $\rm{m^{2}\,s^{-1}}$  & \citet{thiriet_scaling_2019}  \\
Ra$_{\rm crit}^{u}$ & Critical Rayleigh number & 450 & - & \citet{thiriet_scaling_2019}  \\
$a_{\rm rh}$ & Viscosity temperature scale coefficient & 2.44 & - & \citet{thiriet_scaling_2019} \\
$\beta$ & Heat flow scaling exponent & 1/3 & - & \citet{solomatov_scaling_1995} \\
$T_s$ & Surface temperature & 273 & K & \\


\midrule
\multicolumn{5}{c}{\textbf{Variables tested in the parameter study}} \\
$\tau$ & Planet age & 2--4.5, \enspace baseline: 4.5 & Gyr \\
$M_p$ & Planet mass & 0.1--5.0, \enspace baseline: 1.0 & $M_\Earth$ & \citet{rogers_most_2015, zeng_mass-radius_2016} \\
CMF & Core mass fraction & 0--0.4, \enspace baseline: 0.3 & - & \citet{zeng_mass-radius_2016} \\
$\chi_{\rm rad}$ & U and Th budget relative to solar & 0.3--3.0, \enspace baseline: 1.0 & - & \citet{nimmo_radiogenic_2020} \\

\midrule
\multicolumn{5}{c}{\textbf{Unknown random variables}} \\
$E_a$ & Viscosity activation energy & $\mathcal{U}(200, 300)$ & $\rm{kJ\,mol^{-1}}$ & \citet{karato_rheology_1993, zhang_diffusion_2017} \\
$\eta_0$ & Viscosity prefactor & $\mathcal{U}(2.6 \times 10^{10} , 5.3\times 10^{13})$ & Pa s & see section \ref{sec:viscosity-model} in text \\
$A, B, C, D$ & Topography scaling coefficients & $\mathcal{N}(\mathbf{\mu}, \mathbf{\Sigma})$\tablenotemark{a} & - & This work \\

\bottomrule
\end{tabular}
\tablenotetext{a}{\footnotesize with mean $\mathbf{\mu}$ and covariance $\mathbf{\Sigma}$ given by the results of the linear regression (see section \ref{sec:methods-hscaling} and Table \ref{tab:fit}).}
\end{table*}

In a fraction of the CPU time of a full dynamical convection simulation, parameterised convection models can result in similar temperatures to numerical models by tracking heat fluxes across the two thermal boundary layers \citep{thiriet_scaling_2019}. Parameterised convection can also produce a thermal history of the planet, from which we can extract a self-consistent evolution of dynamic topography. Further, such low-cost models invite parameter studies, which naturally we conduct in this segment. Important caveats are discussed in section 4.

We will be exploring how topography changes with planet age, $\tau$, mass, $M_p$, core mass fraction, CMF, and radiogenic heating expressed as an abundance of U and Th relative to the Sun, $\chi_{\rm rad}$. As such, these four parameters are independently and systematically varied between experiments. Meanwhile, we anticipate that some of the biggest uncertainties lie in the unknown mantle rheology. To see how these uncertainties would propagate, rather than testing their effect on $h_{\rm rms}$ explicitly, we will treat the parameters in the viscosity law as uniform random variables. In addition to the viscosity parameters, we also account for model uncertainty by drawing the topography scaling coefficients in (\ref{eq:h_Ra_scaling}) from a multivariate normal distribution whose mean and covariance are given by the results of the regression from section \ref{sec:methods-hscaling}. Table \ref{tab:params} lists all dimensional input parameters used in the 1D model, which the remainder of this section describes.

\subsubsection{Governing energy balances}

The approach outlined here closely follows that of \citet{thiriet_scaling_2019}. The mantle and core temperatures are governed by the 1D energy balances,
\begin{align}\label{eq:T_ODE}
\begin{split}
M_m \, c_{m} \frac{{\rm d}T_m}{{\rm d}t} &= -q_{\rm u} \, A_{\rm u} + q_{\rm rad} \, M_m + q_c \, A_c, \\
M_c \, c_{c} \frac{{\rm d}T_c}{{\rm d}t} &= -q_c \, A_c,
\end{split}
\end{align}
where $t$ is time in s, $M_m$ is the mass of the convecting part of the mantle in kg, $c_{m}$ is the mantle specific heat capacity in ${\rm J\,kg^{-1}\,K^{-1}}$, $q_{\rm rad}$ is the radiogenic heat flux in ${\rm W\,kg^{-1}}$, $q_{u}$ is the heat flux out of the top of the convecting region in ${\rm W\,m^{-2}}$, and $A_{u}$ is the surface area of the top of the convecting region in m$^{2}$. The subscript $u$ denotes the upper boundary layer; the analogous notation with subscript $c$ applies to the core. $M_c$ is found through the core mass fraction. Just as in the 2D models, we explicitly include a mechanical stagnant lid, sitting atop the upper thermal boundary layer, never participating in convection.\footnote{Note that this study does not make a \emph{compositional} distinction (e.g., in density or heat-producing element concentration) between the convecting mantle and the lid. In reality, this mechanical boundary layer would partially overlap with the planetary crust, the latter being the product of bulk mantle that partially melted, generated magmas that rose buoyantly to the surface, and re-crystallised as a lower-density rock.} Our choice of initial conditions for the governing equations are explained in section \ref{sec:methods-lid1D}.

Note also that we assume a perfectly spherical planet. For simplicity, and for consistency with our assumption of incompressibility in the 2D models, we treat $c_m$ and other thermodynamic quantities as constant throughout the mantle (i.e., always equal to their reference values at the top of the convecting mantle); in reality these would vary with the adiabatic profile. This assumption would be a greater source of error for more massive planets with higher pressures at the base of the lithosphere. Although (\ref{eq:T_ODE}) simplifies the problem by omitting other heat fluxes like volcanism (see section \ref{sec:discussion-Ra}), it will suffice in capturing the essential behaviour of a cooling convective planet \citep{jaupart_temperatures_2015}.

\subsubsection{Interior structure}

The radius of the planet, $R_p$, is based on the physically-motivated mass-radius relation in \citet{zeng_mass-radius_2016},
\begin{equation}\label{eq:MR}
\frac{R_p}{R_\Earth} = (1.07 - 0.21\; {\rm CMF})\left(\frac{M_p}{M_\Earth}\right)^{1/3.7},
\end{equation}
whilst the radius of the core, $R_c$, is from \citet{zeng_simple_2017},
\begin{equation}\label{eq:CMF}
R_c = R_p \; {\rm CMF}^{0.5}.
\end{equation}
We use a surface gravity $g_s$ consistent with $M_p$ and $R_p$. Note that Table \ref{tab:params} suggests the mantle density, $\rho_m$, is a constant, but (\ref{eq:MR}) and (\ref{eq:CMF}) assume that density decreases radially outwards such that gravity is constant through the mantle. Our box model can be said to treat $\rho_m$ as a near-surface value, apt for the upper thermal boundary layer typically found at $r \approx 0.99R_p$. Note that (\ref{eq:MR}) and (\ref{eq:CMF}) entail extrapolating equations of state to pressures beyond their validity range, which could lead to errors in $R_p$ and $R_c$, compared to more accurate high-pressure equations of state such as in \citet{hakim_new_2018}. Even at 5 M$_\Earth$, however, the radius predicted by (\ref{eq:MR}) is 1.2\% smaller than that from \citet{hakim_new_2018} for an Earth-like core size. This radius error has no effect on RMS dynamic topography, but decreases ocean basin sizes by 8\%. Significant errors in dynamic topography predictions would come with $R_p$ overinflations of more than 20\%. In detail, accurate mass-radius relations will require tailoring to specific bulk compositions.

In the parameter study, we vary CMF from 0.0 to 0.4, the quoted range for which (\ref{eq:MR}) is valid. Neglecting any potential silicate mass loss after planet differentiation, oxidation chemistry predicts a theoretical upper CMF of 0.34 \citep{dyck_effect_2021}. We consider values of $M_p$ ranging from 0.1 $M_\Earth$ to 5 $M_\Earth$, corresponding to a Mars-sized body and to an equivalent radius slightly below the accustomed upper limit for rocky planets at 1.6 $R_\Earth$ \citep{rogers_most_2015} based on (\ref{eq:MR}) with a CMF of 0.33.

\subsubsection{Mantle rheology}\label{sec:viscosity-model}

The rheology of rocky mantles is thought to obey an Arrhenius law \citep{karato_rheology_1993}. The Arrhenius functional form yields exceedingly large viscosity contrasts over the cold lithosphere---spawning numerical issues in 2D models that preclude its use there. We exploit the Arrhenius form in the 1D model, but to maintain consistency between our 1D and 2D models, we ignore any pressure-dependence and non-Newtonian behaviour. We adopt a canonical law for diffusion creep as a function of temperature,
\begin{equation}
    \label{eq:eta-arrhenius}
\eta(T) = \eta_0 \exp\left(\frac{E_a}{R_b T}\right),
\end{equation}
where $\eta$ is the dynamic viscosity in Pa~s, $R_b = 8.314$ is the gas constant in $\rm{J\,mol^{-1}\,K^{-1}}$, $E_a$ is the activation energy in $\rm{J\,mol^{-1}}$, and $\eta_0$ is a prefactor with the same units as $\eta$. Note that our definition of $\eta_0$ does not act as a ``reference viscosity" sometimes employed; it just encompasses all pre-exponential terms. In natural systems, the mantle viscosity will also depend on pressure; this caveat is discussed in section \ref{sec:discussion-rheology}.

In testing variations of $\eta_0$ and $E_a$, we shall try to capture the uncertainty imparted by unconstrained exoplanet rheologies. Strain rates brought on by the diffusion creep of silicate mantle rock would be strongly affected by both the water content and the bulk mineralogy. For olivine, \citet{karato_rheology_1993} give the canonical wet (water-saturated) and dry (water-free) flow laws: $E_a$ from 240~kJ mol$^{-1}$ in the former to 300~$\rm{kJ\,mol^{-1}}$ in the latter; water weakens the rock. For the pre-exponential coefficient $\eta_0$, the same canonical laws correspond to $1.6 \times 10^{11}$ and $2.6 \times 10^{11}$ Pa~s, which produces a dry olivine viscosity of $\sim$10$^{21}$~Pa~s at 1600~K. 

We also expect to find overall higher viscosities inside planets that have mantles with lower Mg/Si compared to Earth's value of $\sim$1.3 \citep{pagano_chemical_2015, spaargaren_influence_2020, ballmer_diversity_2021}. At Mg/Si \textless~1, the upper mantle composition would be dominated by orthopyroxene; at Mg/Si near 2 it would approach pure olivine. Our coarse treatment considers some empirical end members. We have laws for olivine; \citet{zhang_diffusion_2017} give an Arrhenius flow law for the diffusion creep of enstatite. They find that $E_a = 200$~$\rm{kJ\,mol^{-1}}$, that wet enstatite is approximately 10 times more viscous than wet olivine at depth, and that virtually-dry enstatite is about 100 times more viscous than wet enstatite.


So far, this simple mineralogical paradigm would imply that water-saturated regions of Earth's upper mantle would exhibit the weakest-possible diffusion creep among rocky planets.
To be conservative, we set a minimum $\eta_0$ of $2.6 \times 10^{10}$ Pa~s, an order of magnitude weaker than wet olivine \citep{karato_rheology_1993}. The maximum $\eta_0$ is set at $5.3 \times 10^{13}$~Pa~s, approximating a dry enstatite rheology \citep{zhang_diffusion_2017}. We test $E_a$ between 200~$\rm{kJ\,mol^{-1}}$ and 300~$\rm{kJ\,mol^{-1}}$. Both $E_a$ and $\eta_0$ are drawn from random uniform distributions. By varying these parameters independently, we are likely overestimating the true uncertainty if they are in fact correlated. Note that we do not self-consistently adapt other bulk properties to account for the unknown mineralogy (an invaluable endeavour, but outside the scope of the current manuscript).

\subsubsection{Heat fluxes}

\paragraph{Internal heating} The radiogenic heat flux at $t$ is:
\begin{align}\label{eq:q_rad}
q_{\rm rad} &= \sum_{i = 1}^4 \chi_i c_i h_i \exp\left[\left(\tau - t\right) \frac{\ln 2}{\tau_{1/2, i}}\right], &\\
\chi_i &= \nonumber
\begin{cases}
\chi_{\rm rad} & \textrm{if } i \ge 2 \\
1 & \textrm{otherwise}
\end{cases}
\end{align}
where we are summing over the heat-producing isotopes $^{40}$K, $^{238}$U, $^{235}$U, and $^{232}$Th, $c_i$ is the present-day bulk silicate Earth concentration of the $i^{\rm th}$ isotope in $\rm{kg\,kg^{-1}}$, $h_i$ is the heating contribution in $\rm{W\,kg^{-1}}$, and $\tau_{1/2, i}$ is the half-life in the same units as $t$. Values for these parameters are taken from Table 1 in \citet{oneill_distribution_2020}. Further, for the refractory elements U and Th, we multiply the summand by a common factor $\chi_{\rm rad}$ to reflect potentially-extraterrestrial variations in the abundances of these $r$-process elements. As surveyed in \citet{nimmo_radiogenic_2020}, U and Th abundances are conservatively expected to vary across Sun-like stars from between 30\% to 300\% of the solar value, which---assuming that relative mantle concentrations directly reflect relative stellar abundances \citep{thiabaud_elemental_2015, hinkel_starplanet_2018, putirka_composition_2019, adibekyan_chemical_2021}---translates to a range in $q_{\rm rad}$ of 2.22--14.34 $\rm{pW\,kg^{-1}}$ at 4.5 Gyr, with the baseline value equivalent to $5.36 \times 10^{-12}\,\rm{W\,kg^{-1}}$. (We ignore the unconstrained variations in $^{40}$K, a volatile isotope which in any case contributes less heating with age than refractory U and Th.) Although we do not account for the galactic chemical evolution of U and Th abundances as a function of stellar age \citep{frank_radiogenic_2014}, some of this variation is captured in $\chi_{\rm rad}$ regardless. 

\paragraph{Thermal boundary layers} Across the upper and lower thermal boundary layers, heat fluxes are conductive:
\begin{equation}
    q_{u, c} = k_m \frac{\Delta T^{u, c}}{\delta^{u, c}_{\rm rh}}, \label{eq:q_u}
\end{equation}
where $k_m$ is the mantle thermal conductivity in $\rm{W\,m^{-1}\,K^{-1}}$, $\Delta T^u$ (respectively $\Delta T^c$) is the temperature contrast across the upper (lower) boundary layer in K, and $\delta^{u}_{\rm rh}$ ($\delta^{c}_{\rm rh}$) the thickness in m. 

The thermal boundary layer thicknesses are controlled by their local Rayleigh numbers:
\begin{align}
\delta^{u, c}_{\rm rh} &= (R_{\rm lid} - R_c) \left(\frac{{\rm Ra}^{u, c}_{{\rm crit}}}{{\rm Ra}^{u, c}_{{\rm rh}}}\right)^\beta,  \label{eq:d_u}\\ 
{\rm Ra}^{u, c}_{{\rm rh}} &= \frac{\alpha\rho g^{u, c} \Delta T^{u, c} (R_{\rm lid} - R_c)^3}{\kappa \eta(T^{u, c})}, \label{eq:Ra_rh}
\end{align}
where Ra$^{u, c}_{{\rm rh}}$ is the local Rayleigh number, Ra$^u_{{\rm crit}}$ is the critical Rayleigh number for convection, and $\beta$ is a constant which can be obtained from either experiments or theory. For both thermal boundary layers we take $\beta = 1/3$, such that $q_u$ is independent of $d$; the boundary layers are assumed to be in a state of marginal stability \citep[e.g.,][]{solomatov_scaling_1995}. The value of $\beta$ is tied physically to the planet's dominant cooling mechanism, which strongly depends on the tectonic mode \citep{lenardic_diversity_2018, seales_uncertainty_2020}. The choice made here is appropriate for chaotically-time dependent, stagnant lid convection with temperature-dependent viscosity \citep{solomatov_scaling_1995, solomatov_scaling_2000}. Other fitting choices do not significantly change our results \citep{thiriet_scaling_2019}.  

For the upper thermal boundary layer, we have: $\Delta T^u = T_m - T_{\rm lid}$; $\eta(T^u) = \eta(T_m)$; $g^u = g_s$; and we fix Ra$^u_{{\rm crit}}$ at 450. Now for the lower layer, this becomes: $\Delta T^c = T_c - T_m$; $\eta(T^c) = \eta[(T_c + T_m)/2]$; $g^c$ the gravity at $R_c$; and after \citet{deschamps_inversion_2000}, Ra$_{{\rm crit}, c}$ = 0.28Ra$_i^{0.21}$, with Ra$_i$ the interior Rayleigh number defined for 1D convection in (\ref{eq:Ra_i_1D}). Although Ra$_{{\rm crit}, c}$ can be tricky to parameterise, $T_c$ tends to equilibriate with $T_m$ fairly quickly under this setup, hence $q_c \ll q_u$.

Finally, the temperature $T_{\rm lid}$ at the base of the lid in K (identically, at the top of the convecting region) is obtained for parameterised convection in a similar way to numerical models. The temperature drop between $T_m$ and $T_{\rm lid}$ is proportional to the so-called viscous temperature scale, $\Delta T_\nu$ \citep{davaille_transient_1993}:
\begin{align}
\label{eq:Tl}
T_{\rm lid} &= T_m - \Delta T_{\rm rh} = T_m - a_{\rm rh} \Delta T_{v},\\
\label{eq:T_rh}
\Delta T_\nu &= \frac{\eta(T_m)}{{\rm d}\eta/{\rm d}T\vert_{T_m}} = \frac{R_b T_m^2}{E_a}.
\end{align}
The coefficient $a_{\rm rh}$ is empirically-determined; we adopt a value of 2.44 for $\beta = 1/3$ based on \citeauthor{thiriet_scaling_2019}'s (\citeyear{thiriet_scaling_2019}) fits to 3D spherical convection simulations. The radius $R_{\rm lid}$ of this temperature coordinate is described in the next section.

\subsubsection{Stagnant lid thickness and the final governing equation}\label{sec:methods-lid1D}

The lid does not instantly grow or shrink in response to a change in the heat flux coming from the upper thermal boundary layer. Rather, there is a lag in which $\delta_{\rm lid}$ adjusts such that the difference between the flux out of the top of the lid and the flux into the base of the lid is minimised:
\begin{equation}\label{eq:D_l}
\frac{{\rm d} \delta_{\rm lid}}{{\rm d} t} = \frac{q_{\rm lid}\vert_{R_{\rm lid}} -q_u}{\rho_m c_{m} (T_m - T_{\rm lid})},
\end{equation}
where the heat flux profile of the lid, $q_{\rm lid}(r)$ in ${\rm W\,m^{-2}}$, is obtained by solving the steady-state conductive heat transfer equation in spherical geometry with boundary conditions ($R_{\rm lid}, T_{\rm lid}$) and ($R_p, T_s$) where $T_s$ is the surface temperature in K, and with internal heating equal to the mantle $q_{\rm rad}$ (in reality, we might anticipate higher concentrations of lithophiles U, Th, and K in the lid). This steady-state formulation ignores the time-dependence of heat conduction in the lid, leading to errors compared to a time-dependent model in the surface heat flux of $\lesssim 5\,\rm{mW\,m^{-2}}$ for a Mars-sized planet. A smaller error is expected for larger planets with thinner lids \citep{thiriet_scaling_2019}. 

We account for the mass of the convecting region changing with $\delta_{\rm lid}$ by subtracting the lid mass, $\rho_m 4\pi/3  (R_p^3 - R_{\rm lid}^3)$, from the fixed quantity $M_p(1 - \text{CMF})$. At each time step we also update $R_{\rm lid} = R_p - \delta_{\rm lid}$. Thus (\ref{eq:D_l}) presents a third differential equation that must be solved simultaneously with (\ref{eq:T_ODE}). We solve this system of equations using the explicit Runge-Kutta method of order 5. The initial conditions, $T_{m,0}$, $T_{c,0}$, and $\delta_{{\rm lid}, 0}$ reflect the unknown formation history of the planet---the leftover gravitational energy of accretion and core segregation, and the crystallisation of the primordial magma ocean(s). To bypass this uncertainty, we only consider simulations that have converged to a memoryless state. That is, we prime each experiment by running it forwards from $t$ = $-$5 to 0~Gyr, and using the solution at 0~Gyr as the initial conditions. Then (\ref{eq:T_ODE}) and (\ref{eq:D_l}) are solved again from $t$ = 0 to $\tau$.

\subsubsection{Dynamic topography}

Once we have a solution for the planet's thermal history, we combine these results with (\ref{eq:h_Ra_scaling}) to find $h_{\rm  rms}^\prime$. Since we have $b$ and Ra$_i$ forming the basis of the topography scaling from 2D experiments, applying (\ref{eq:h_Ra_scaling}) to 1D thermal histories requires writing 1D-appropriate analogues of these two variables. An analogue of Ra$_i$ is quite straightforward; for parameterised convection this variable is defined \textit{a posteriori} as
\begin{equation}\label{eq:Ra_i_1D}
    {\rm Ra}_i = \frac{\alpha_m\rho_m g_s \Delta T \left(R_p - R_c\right)^3}{\kappa_m \eta(T_m)},
\end{equation}
where $\Delta T = T_s - T_c$. This equation is the same as (\ref{eq:Ra_i}) using the dimensional parameters for the mantle in Table \ref{tab:params} and simply letting the interior viscosity $\eta(T_i) = \eta(T_m)$. For our runs, $T_c \approx T_m$. Note also that Ra$_i$ differs from Ra$_{\rm rh}^u$ (\ref{eq:Ra_rh}) in that the latter excludes the stagnant lid from its domain.

Meanwhile, $b$ as defined in the exponential viscosity law must be related to Arrhenius law parameters, since the 1D convection model the latter, more-realistic law. \citet{moresi_numerical_1995} demonstrate such an exponential approximation to an Arrhenius law. The approximation comes from the idea that in the stagnant lid regime, it is the local rheological gradient over the upper thermal boundary layer that propels temperature-dependent viscosity convection, rather than the total domain viscosity contrast, $\Delta \eta$ \citep{davaille_transient_1993}. One can therefore write $\eta(T) \sim \exp\left[\left( \Delta T / \Delta T_\nu \right) T \right]$, where the viscous temperature scale $\Delta T_\nu$ is re-scaled by $\Delta T$ to make the temperature prefactor dimensionless. From (\ref{eq:T_rh}) this implies
\begin{equation} \label{eq:b-1D}
    b = \frac{\Delta T}{R_b T_m^2 / E_a}.
\end{equation}
In 2D applications, setting $T_m$ at the interior temperature just below the upper thermal boundary layer would create a viscosity profile which is most closely aligned to the Arrhenius profile, especially over the key region of the upper thermal boundary layer \citep{moresi_numerical_1995}.

Finally, the dimensionless $h_{\rm rms}^\prime$ resulting from (\ref{eq:b-1D}), (\ref{eq:Ra_i_1D}), and (\ref{eq:h_Ra_scaling}) is scaled by $\alpha_m \Delta T d$ (\ref{eq:dimensionalise}) to get the dimensional $h_{\rm rms}$. To clarify, we do consider the whole domain in the dimensionalisation, so $d = R_p - R_c$ and again $\Delta T = T_c - T_s$; the fact that several of these constituents evolve with time means that $h_{\rm rms}$ is a function of the age of the planet.

These calculations so far have assumed subaerial topography. Water-loaded topography would be higher by a factor of $\rho_m/(\rho_m - \rho_w) \approx 1.5$, where $\rho_w$ is the density of water.

It is worth mentioning at this point that the dependence in several places on $T_s$---inside the definition of $b$ in particular---means there is a certain sensitivity of $h_{\rm rms}$ to this free parameter. For example, all else held constant at the baseline value (Table \ref{tab:params}), increasing $T_s$ from 273 to 373 K is associated with a 30\% decrease in $h_{\rm rms}$. However, because this study is only concerned with temperate planets which have a narrow range in $T_s$, we do not consider its effect on topography.


\subsection{Expansion to maps and the volume of ocean basins}

We have based our scaling relationship on $h_{\rm rms}$ (section \ref{sec:methods-hscaling}), yet it is the peak topography, $h_{\rm peak}$, that controls how much water a planet's surface reservoirs can hold at the maximum capacity. Therefore we require the peak topography associated with an RMS value in a 3D spherical geometry, given assumptions about topography's distribution. 

Appendix \ref{sec:sph-harms} explains the relevant spherical harmonics method in more detail. Suppose we have a log-linear power spectrum, which fiducially describes dynamic topography amplitudes on a sphere. Essentially, for each run of 1D thermal evolution, we transpose the power spectrum vertically such that its frequency-domain RMS value matches the spatial-domain RMS value expected from the $h_{\rm rms}({\rm Ra}_i, b)$ scaling function. The transposed spectrum is expanded onto a 2D map, $h(x, y)$, which has its own $h_{\rm peak} = {\rm max}(h)$. The volumetric ocean basin capacity in cubic metres---the main intended application of our topography modelling---is estimated as
\begin{align}\begin{split}\label{eq:ocean-integral}
V_{\rm cap} &= \frac{\rho_m}{\rho_m - \rho_w} \int \left[h_{\rm peak} - h(x, y)\right] \, {\rm d} S \\
&= \frac{\rho_m}{\rho_m - \rho_w} 4 \pi R_p^2 h_{\rm peak},
\end{split}
\end{align}
where the integral is over the surface $S$ and the 2D map is multiplied by the density ratio term to account for water-loaded topography (our purpose here entails that the whole map is underwater, save for the single grid point corresponding to $h_{\rm peak}$). The actual basin capacities of Venus, Earth, and Mars defined this way are 3.4, 3.3, and 2.9 Earth oceans respectively---we expect to find lower values by considering only dynamic topography.

Robust models of dynamic topography power spectra are not available at this time. Instead, for the spectrum needed above, we explore three hypothetical scenarios. The first and most simple model is that all topography behaves like red noise, as per the historical paradigm introduced in section \ref{sec:intro-spectrum} \citep[e.g.,][]{turcotte_fractal_1987}. The second option is to represent empirical dynamic topography with the observed shape of Venus---although broad regions of Venus' highlands indicate isostatic support, so the resulting spectral distribution should reflect a mix of support mechanisms \citep[e.g.,][]{kiefer_dynamic_1986, arkanihamed_analysis_1996, simons_localization_1997, yang_separation_2016}; further, Venus may not be a perfectly archetypal stagnant lid planet, and be better described instead by a plutonic-squishy lid regime \citep{lourenco_plutonic-squishy_2020}. Option three is to be consistent with the pure dynamic topography we already produced to feed our scaling functions: we derive time-averaged power spectral densities from the numerical topography profiles, to which we fit a generic model.

Although the present study only considers dynamic topography, this same framework could be applied to any kind of topography on a planet as long as we can infer its spectral distribution.

\section{Results} \label{sec:results}

\subsection{Numerical modelling results}

\begin{figure*}
\epsscale{1.1}
\plotone{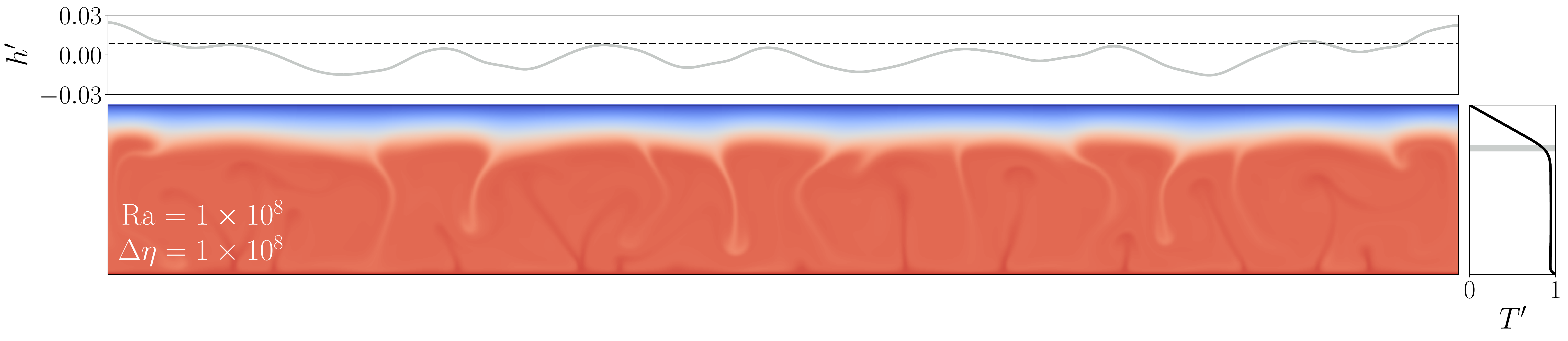}

    \caption{Snapshot from a single time step of the dimensionless temperature field \textit{(bottom)}, surface dynamic topography, $h^\prime$ \textit{(top)}, and temperature profile, $T^\prime$ \textit{(right)}, for chaotically time-dependent convection in the stagnant lid regime. This example shows Ra$_1 = 1 \times 10^8$ and $\Delta \eta = 1 \times 10^8$. The dimensionless temperatures range from 0 (cold; blue) to 1 (hot; red). The grey box in the temperature profile shows the instantaneous location and thickness of the upper thermal boundary layer. The vertical scale of $h^\prime$ is exaggerated. 
    \label{fig:aspect-output}}
\end{figure*}

The products of numerically-modelled chaotic stagnant lid convection include time-dependent, dimensionless temperature fields and surface dynamic topography profiles (figure \ref{fig:aspect-output}). For each case, temporally- and horizontally-averaged temperature fields are used to calculate $T_i^\prime$, Ra$_i$, and other convective parameters; full outputs can be found in Table \ref{tab:aspect-out} in the appendix to this paper. Average $T^\prime$ profiles hardly vary in time, hence neither does $T_i^\prime$ nor the average position of the upper thermal boundary layer's base. Stepping up Ra$_1$ thins $\delta_{\rm rh}^\prime$, and lowers the RMS height of topography in the regime we explore numerically. Increasing $\Delta \eta$ thickens the stagnant lid because high viscosities are reached at lower depths; this is also associated with a slight increase in $\delta_{\rm rh}^\prime$.

\subsubsection{Fit to RMS height of topography} \label{sec:results-scaling}

\begin{table}
\centering
\caption{Topography scaling coefficients and their errors obtained from fitting a multiple linear regression model with an interaction term to equation (\ref{eq:h_Ra_scaling}). The bottom row reports the residual variance, $\sigma^2_{\rm res}$, of the fit. \label{tab:fit}}
\footnotesize
\begin{tabular}{@{} r r r r r  @{}}
\toprule
& $A$ & $B$ & $C$ & $D$ \\
\midrule
Best fit & 9.581 & -0.5818 & -1.510 & 0.07536 \\
Standard deviation & 3.298 & 0.1859 & 0.4220 & 0.02379 \\
\midrule
\multicolumn{5}{r}{$\sigma^2_{\rm res} = 1.584 \times 10^{-3}$} \\
\end{tabular}
\end{table}

\begin{figure}
    \centering
    \epsscale{1.1}
    \plotone{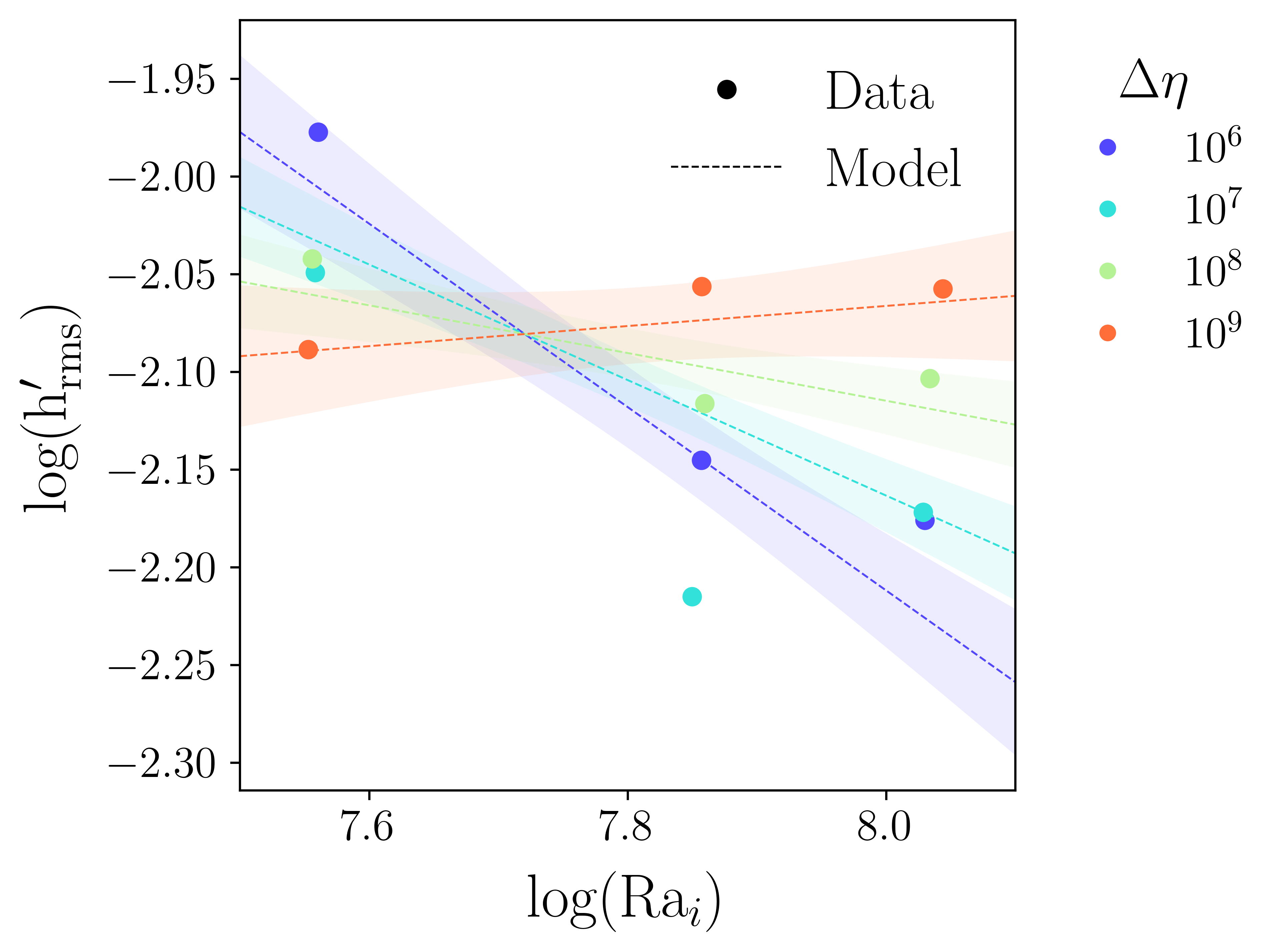}
    \caption{Fitted scaling relationship for dimensionless RMS dynamic topography, $h^\prime_{\rm rms}$, from 2D numerical convection simulations ($n=12$). Topography is given by a four-parameter linear model, which depends on the interior Rayleigh number, Ra$_i$, and the viscosity temperature prefactor, $b = \ln(\Delta \eta)$. Markers represent individual cases (see Table \ref{tab:aspect}) and are coloured according to $\Delta \eta$. The uncertainties on $h^\prime_{\rm rms}$, taken to be the standard errors of the mean, are smaller than the marker size. Dashed lines represent the best-fit parameter combination at discrete $\ln(\Delta \eta)$. Swaths span one standard deviation of the response variable, propagated from the covariance matrix of the fit.}
    \label{fig:2D-h-scale}
\end{figure}

Figure \ref{fig:2D-h-scale} shows the four-parameter linear fit between log($h_{\rm rms}^\prime$), log(Ra$_i$), and $\ln(\Delta \eta)$, using the functional form in (\ref{eq:h_Ra_scaling}). Best-fit parameter values and standard deviations are given in Table \ref{tab:fit}. The residual variance of this fit is $\sigma^2_{\rm res} \sim 10^{-3}$, equal to the sum of squares error divided by the degrees of freedom. Because the fitted data correspond to averages over model time, the standard errors of the mean independent and dependent variables are all small and do not impact the regression.

The key piece of information from this section is that chaotic convection with temperature-dependent viscosity does not lend itself to constant power-law scalings of $h_{\rm rms}^\prime$ with Ra$_i$ (or Ra$_1$). The value of $\Delta \eta$ is effectively altering the slope of $\log(h_{\rm rms}^\prime)$ with $\log({\rm Ra}_i)$. Smaller viscosity contrasts of $10^7$ ($b=16$) and below are associated with strongly negative slopes. With increasing $\Delta \eta$, the slope grows systematically shallower, until it changes sign between $\Delta \eta = 10^8$ ($b=18$) and $\Delta \eta = 10^9$ ($b=20$). 
Conversely, the effect of Ra$_i$ on $\Delta \eta$ is such that at higher Ra$_i$ above $\sim 6 \times 10^7$, large viscosity contrasts favour high RMS topography, whilst at lower Ra$_i$ below $\sim 6 \times 10^7$, small viscosity contrasts favour high RMS topography. At Ra$_i \sim 6 \times 10^7$, these slopes ``cross over" and the effect of $\Delta \eta$ disappears. 

Evidently this behaviour is governed by a complex, chaotic system; extracting a general mechanistic understanding is compromised by the limited number of runs performed here. The effect of $\Delta \eta$ to increase $h_{\rm rms}^\prime$ may be related to thermal isostatic uplift within the stagnant lid \citep{kucinskas_isostatic_1994, moore_lithospheric_1995, orth_isostatic_2011}. We  include thermal isostasy as part of the full dynamic topography. Under a swell, hot low-density upwelling material extends to shallower depths. To compensate, the cold, dense overlying lithosphere grows thinner, and it is buoyed upwards. It can be shown that the maximum amount of thinning is directly proportional to the average lithospheric thickness. Hence, higher-viscosity-contrast convection, with its deeper lid bases, will enable a greater magnitude of thermal thinning. Meanwhile, smaller Ra$_i$ are associated with thicker $\delta_{\rm rh}$, to which dynamic topography should be proportional \citep{parsons_relationship_1983}. (For a constant $\Delta \eta$, lowering Ra$_1$ also slightly increases $\delta_{\rm lid}$ and thus the potential for thermal thinning.) We speculate that there is a trade-off whereby the $\Delta \eta$ effect dominates when stagnant lids are already thick and when convection is too vigorous to support high topography in its thin thermal boundary layers. Conversely, for lids that are not particularly thick, Ra$_i$ (and $\delta_{\rm rh}$) become more relevant.

A corollary of this is that at the still-higher values of Ra$_i$ expected for realistic rocky planets (up to several orders of magnitude beyond the range amenable to numerics; see discussion in section \ref{sec:discussion-extrap}), the sensitivity of $h_{\rm rms}^\prime$ to the viscosity scale becomes quite high indeed. If the absolute viscosity follows an exponential law, $\eta(T) \sim \exp(-b T)$, high $b$ is associated with low $\eta$ for the same $T$, implying low $h_{\rm rms}^\prime$.

\subsection{Parameterised modelling results}



\subsubsection{Thermal evolution}

\begin{figure}
    \centering
    \epsscale{1.15}
    \plotone{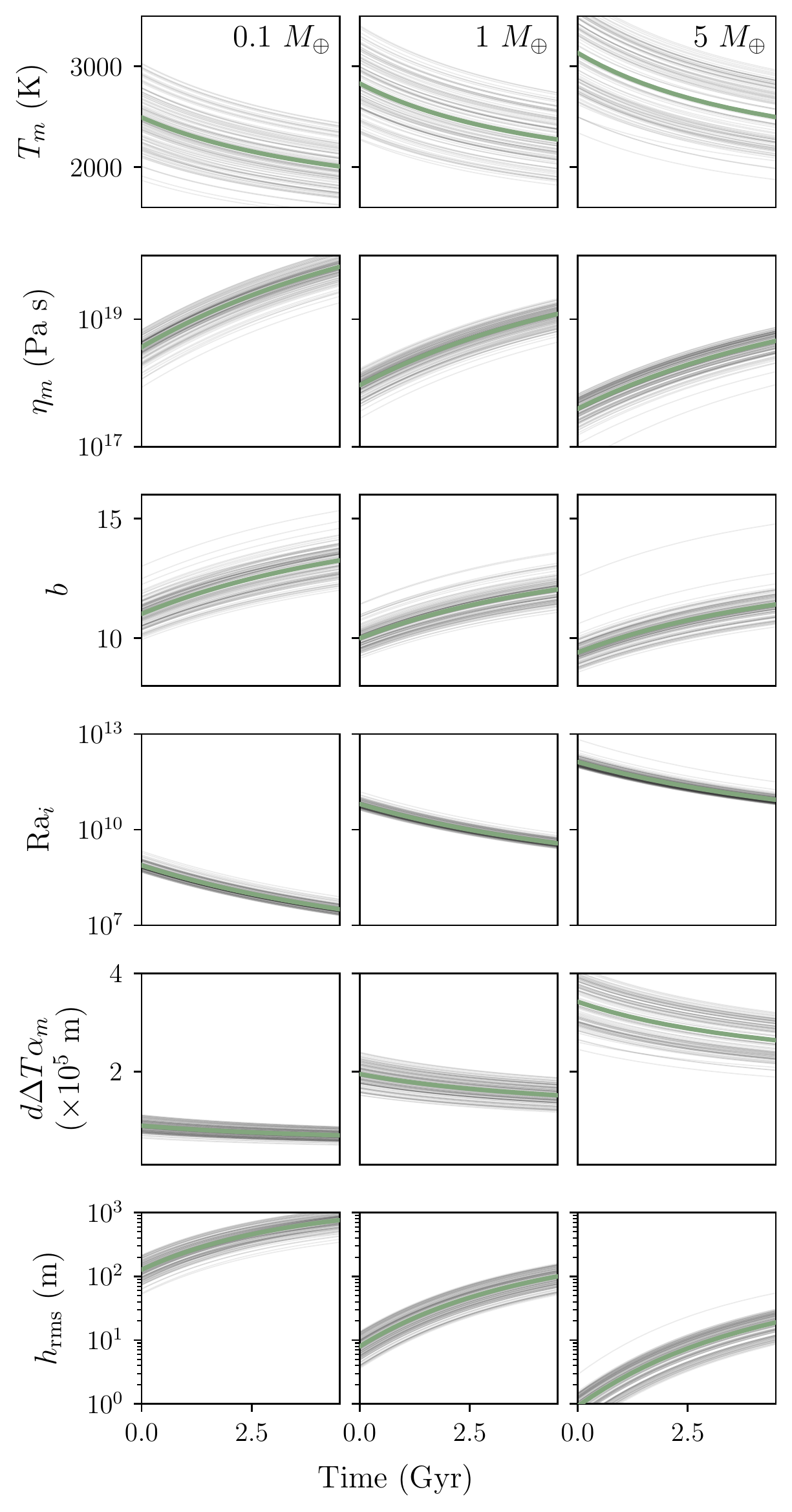}
    \caption{Thermal evolutions sampled from the 1D model ensemble, as a function of time in Gyr. From top to bottom: mantle temperature, $T_m$ in K, mantle viscosity, $\eta_m$ in Pa s, dimensionless inverse viscous temperature scale, $b$, interior Rayleigh number, Ra$_i$, topography dimensionalistion factor, $d \Delta T \alpha_m$ in m, and RMS dynamic topography, $h_{\rm rms}$ in m. Columns compare planet masses from 0.1 $M_\Earth$ \textit{(left)}, through 1 $M_\Earth$ \textit{(centre)}, to 5 $M_\Earth$ \textit{(right)}. Each thin black line ($n = 500$) represents a single evolution, drawing random values of the unknown viscosity activation energy and prefactor, hence an evolutionary spread. Green lines follow the ensemble mean (for Ra$_i$, which is log-normally distributed, this is the log-normal mean). All runs use baseline values of the core mass fraction and radioisotope budget. Parameter values and random variable distributions are given in Table \ref{tab:params}.
    }
    \label{fig:1D-evolution}
\end{figure}

Underlying thermal histories are sampled in figure \ref{fig:1D-evolution}. Because all test planets are initialised at quasi-equilibriated temperatures and stagnant lid thickness, their evolutionary paths reflect secular cooling alone, which track roughly parallel at around $-100$~K Gyr$^{-1}$. Radiogenic heating inevitably declines with age, with surface heat losses lagging behind slightly; the present-day Urey ratios are $\sim$0.65 depending on planet mass. 

Interior temperatures and Ra$_i$ increase with $M_p$ as anticipated from simple scaling laws. We expect the heat flux $q_u$ to increase linearly with planet radius for a fixed internal heat generation rate. This implies that $q_u \propto M_p^{1/3}$, ignoring compression. We can rewrite (\ref{eq:q_u})--(\ref{eq:T_rh}) as
\begin{equation}
    \eta_u = \frac{\rho_m g^u \alpha_m k_m^3 a^4_{\rm rh} \Delta T_\nu^4}{\kappa_m q_u^3 {\rm Ra}^u_{\rm crit}},
\end{equation}
Thus we have $\eta_u \propto M_p^{-1}$; (\ref{eq:Ra_i_1D}) leads to Ra$_i \propto M_p^2$ for approximately the same temperature difference. A five-times more massive planet has a 25-times larger Ra$_i$ \citep[see also][]{stevenson_les_2003,  kite_geodynamics_2009}.

\begin{figure*}
    \centering
    \epsscale{1.17}
    \plotone{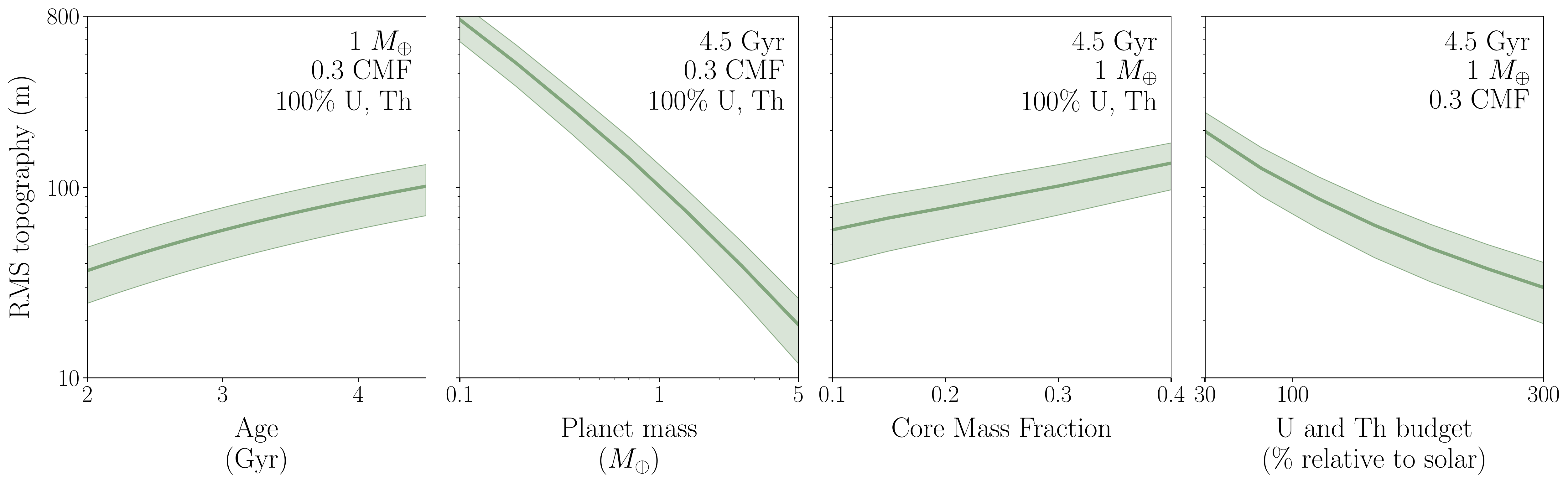}
    \caption{Variations of the RMS dynamic topography based on 1D thermal histories, as a function of select exoplanetary properties which might be constrained in the future: from left to right these are the planet age in Gyr, mass in $M_\oplus$, core mass fraction, and abundance of radioactive U and Th relative to the solar value. Solid lines show the ensemble mean of 10,000 test planets with uniformly-random viscosity activation energies and prefactors, and with normally-random topography scaling coefficients. Swaths span one standard deviation from the mean. These calculations show subaerial topography; water-loaded topography would be $\sim$1.5$\times$ higher. Parameter values and random variable distributions are given in Table \ref{tab:params}.
    }
    \label{fig:1D-h-scale}
\end{figure*}


Figure \ref{fig:1D-evolution} illustrates how uncertainty in the viscosity law parameters $E_a$ and $\eta_0$ affects the spread and mean behaviour of the dimensional $h_{\rm rms}$ and its physical constituents over time. Temperature-dependent viscosity exhibits self-regulating behaviour: a slight increase in temperature lowers the viscosity, hence more vigorous convection via (\ref{eq:Ra}). This leads to more efficient heat loss out of the top of the convecting cell, lowering temperatures in turn. This positive feedback is not visible in a single run (which are already at quasi-steady-state in our case), but we do see the effect at play over the entire ensemble: its range of $\eta_m(t)$ is always less than an order of magnitude, despite a three-order range in $\eta_0$. Meanwhile, $T_m$ is adjusting such that $q_u$ approaches a balanced state for a given $q_{\rm rad}$ and surface area-to-volume ratio. Hence the rheological uncertainty manifests itself in $T_m$. 

We note that these calculated Ra$_i$ values are on average higher for a given $M_p$ than those commonly associated with Venus or Mars. The thermal Rayleigh numbers of real planets require some dexterity to extract, but the few constraints available suggest a value on the order of $10^6$ for Mars \citep{kiefer_melting_2003, samuel_rheology_2019}. Constraints for Venus are even more scarce, but previous work employs Ra at upper mantle temperatures on the order of $10^7$ up to $10^8$ \citep{huang_constraints_2013, king_venus_2018}. This discrepancy is partly explained by the more viscous mantles we permit in this exoplanet study. Further caveats to our Ra$_i$ estimates are discussed in section \ref{sec:discussion-Ra}.

The dimensional $h_{\rm rms}$ reflects a trade-off between $b$, Ra$_i$, and the dimensionalisation factor $d \Delta T \alpha_m$ through (\ref{eq:dimensionalise}) and (\ref{eq:h_Ra_scaling}). Extrapolating figure \ref{fig:2D-h-scale} would imply that, in the $b$-Ra$_i$ regime of the 1D models, high $h_{\rm rms}$ is favoured with high $b$ and low Ra$_i$. Thus deep, hot, weak mantles are doubly-inhibited from having any remarkable topography. It is clear from figure \ref{fig:1D-evolution} that deeper mantles are not enough to make up for lost $h_{\rm rms}^\prime$. 

Ultimately, the thermal state plays a main role in limiting the amplitude of dynamic topography. Hotter mantles necessitate lower viscosities, more vigorous convection, and thinner thermal boundary layers. Within these thinner boundary layers, there may be less scope for density variations related to thermal expansion. If we know some property of a planet to have a strong effect on its interior temperatures, then we might expect it to also impact its dynamic topography.

\subsubsection{Dynamic topography as a function of bulk exoplanetary properties}
\label{sec:results-parameters}

We now test the topographic reaction to planet age, mass, CMF, and radioisotope budget (figure \ref{fig:1D-h-scale}). We find $h_{\rm rms}$ to decrease with $M_p$ and $\chi_{\rm rad}$, and increase with age and CMF. Assuming that the $x$-axes in this figure cover the limits within which we expect to find most rocky exoplanets, then it is plausible that the resulting $y$ range marks the variability of pure dynamic topography which nature could manifest, if our scaling relationship indeed applies. The fact that $h_{\rm rms}$ drops by the largest absolute amounts over $M_p$ and $\chi_{\rm rad}$ reflects the geodynamic significance of these parameters, as well as the spread over which we would expect to find rocky planets. The senses of change of $h_{\rm rms}$ with $M_p$ and $\chi_{\rm rad}$ are predictable from their known effects on $T_m$. That is, hotter interiors are expected for massive, U- and Th-rich planets, hence lower $h_{\rm rms}$. Uncertainty in $h_{\rm rms}$ predictions is tied to uncertainty around the underlying thermal histories: yet another clue to the immeasurable usefulness of characterising this uncertainty more rigorously \citep[e.g.,][]{seales_uncertainty_2020}.

The raw values of $h_{\rm rms}$ predicted by our scaling relationship are on the order of hundreds of metres, whilst the hottest planets can exhibit mere tens of metres of dynamic topography. In fact, due to inherent self-regulation, it is difficult to achieve significantly higher topographies in our 1D model while keeping to Earth-like values of the free parameters. This result may seem very low when compared to the heights of typical topographic features seen across the Solar System. However, a fair comparison requires isolating an RMS height of just the dynamic component of topography; this is not model-independent, as we will discuss (section \ref{sec:benchmarks}).

\subsection{Ocean basin capacity scalings}
\label{sec:results-ocean}

\begin{figure*}
    \centering
    \epsscale{1.15}
    \plotone{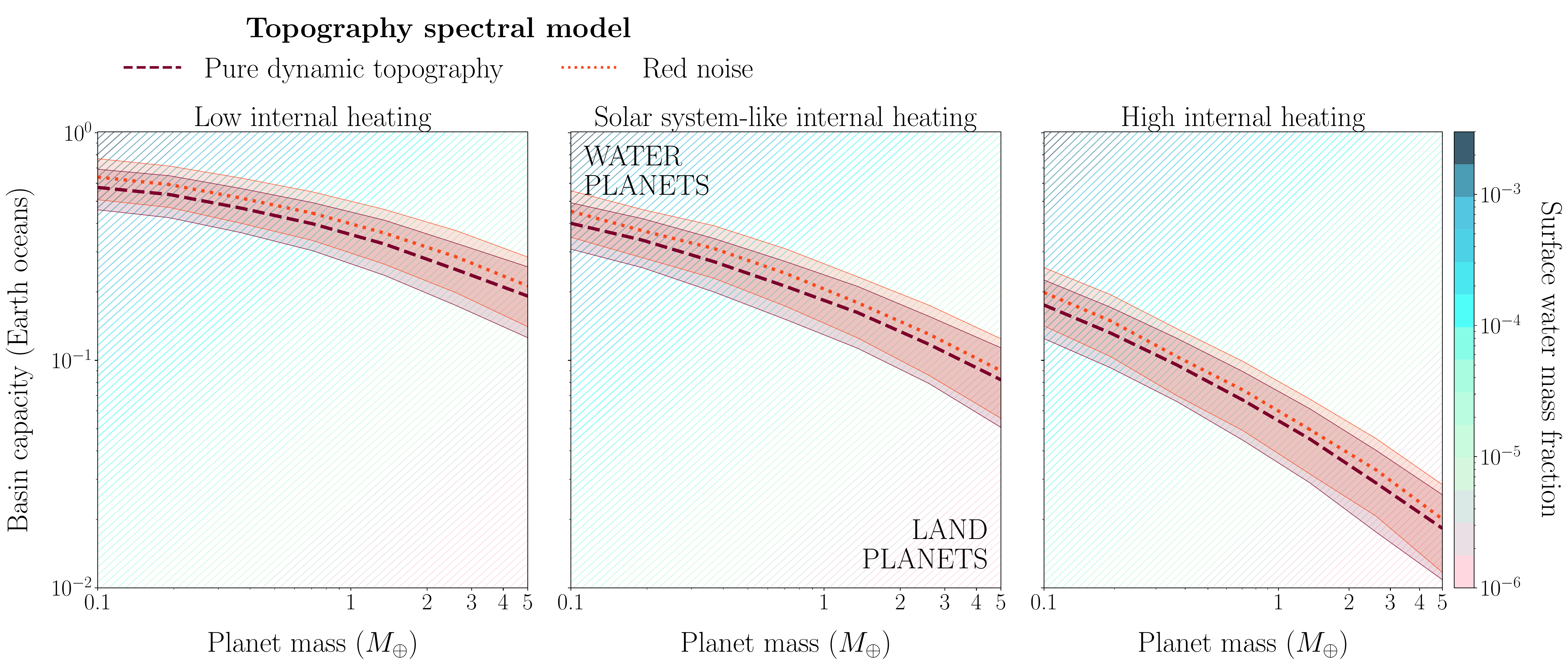}
    \caption{The competition between the ocean basin capacity and the surface water budget with increasing planet mass, expressed in terms of Earth ocean volumes. The three panels are based on, from left to right, the 1st percentile value, solar value, and 99th percentile value of expected mantle heat production across rocky exoplanets \citep{nimmo_radiogenic_2020}. Basin capacities as a function of planet mass are calculated from either the pure dynamic topography spectral model (dashed purple lines) or the red noise model (dotted red lines). The empirical Venus model overlaps the pure dynamic topography and is not shown. Random rheological parameters propagate through the model; the resulting 1$\sigma$ variation is represented by the swaths. In the background, the solid contours follow lines of constant surface water as a fraction of planet mass (note a modern Earth value of $\sim$200 ppm). Thermal histories refer to a 4.5-Gyr planet with a core mass fraction of 0.33.}
    \label{fig:ocn-vol}
\end{figure*}

We have tested three fiducial spectral models to find a relationship between the RMS and peak value of dynamic topography. The theoretical red noise model, the empirical Venus model, and the numerical dynamic topography model all produce an $h_{\rm peak}$ which is, on average, some constant scalar multiple of $h_{\rm rms}$. For both numerical dynamic topography and the total Venus topography, $h_{\rm peak} \approx 3.5 h_{\rm rms}$, and for red noise topography, $h_{\rm peak} \approx 3.9 h_{\rm rms}$. (For a pink noise structure similar to Earth's observed dynamic topography, $h_{\rm peak} \approx 4.0 h_{\rm rms}$.) 

We use our $h_{\rm peak}$ estimations to derive the ocean basin volume capacity $V_{\rm cap}$ as a function of planet mass (figure \ref{fig:ocn-vol}). This quantity represents the smallest volume of surface liquid water that would entirely inundate a planet. The actual land fraction requires knowing the ocean mass. We leave sea level as an unknown quantity and simply consider fiducial surface water budget scenarios. Specifically, we treat the amount of surface water as a constant mass fraction of $M_p$. This parameterisation brackets the planet's total water budget with its volatile partitioning between the interior and exterior---in reality the amount of water stored in the mantle would affect the planet's thermal evolution through its rheology (and melting history, which is not modelled). 

As the basin volume capacity changes with $M_p$, so too does the water volume corresponding to this mass fraction (we assume a density of $1000\,\rm{kg\,m^{-3}}$; salt water is slightly denser). Figure \ref{fig:ocn-vol} can be read as follows: for a given surface water budget, the planet mass where this contour intersects the basin capacity gives the most massive planet that could sustain land with dynamic topography alone. For example, a 1-$M_\Earth$, 4.5-Gyr-old planet endowed with solar U and Th could hold about 0.3 Earth oceans on its surface. The internal heating rate has a strong influence on $V_{\rm cap}$.

Figure \ref{fig:ocn-vol} compares different assumptions about the spectral distribution of topography, which would affect the relationship between the peak and RMS topography. The dynamic topography and Venus models overlap identically, and the red noise spectrum results in only slightly larger $V_{\rm cap}$, seemingly because they are very similar in the low-degree regions where most of their power is concentrated. The basin volume corresponding to an infinitesimally-small but nonzero land area is insensitive to the distribution of topography at high frequencies.

We can formulate these results in terms of a simple scaling analysis. Equation (\ref{eq:ocean-integral}) can be written as $M_\text{cap} = 4 \pi R_p^2 \rho_w  \rho_m / (\rho_m - \rho_w) h_{\text{peak}}$, where $M_\text{cap}$ is the ocean basin capacity in kg. For Earth's ocean mass ($1.4 \times 10^{21}$~kg), this means a peak topography $h_{\text{peak}}$ less than 2.7 km leads to a waterworld. If $h_{\text{peak}}$ were independent of planet mass, we would expect $M_\text{cap} \propto M_p^{2/3}$ due to the increase in surface area alone (the mass-radius relation in (\ref{eq:MR}) gives a slightly shallower power due to compression). However, we have $h_{\text{peak}}$ strongly decreasing with increasing mass. For dry olivine and solar U and Th abundances, $h_{\text{peak}} \propto h_{\text{rms}} \propto M_p^{-0.5}$. From (\ref{eq:ocean-integral}), $M_\text{cap} \propto R_p^2 h_{\rm peak}$, so $M_\text{cap} \propto M_p^{0.04}$ using (\ref{eq:MR}). Warmer, less viscous interiors decrease this exponent, so the most massive rocky planets have the smallest basin capacities even though they have the largest surface areas. If the pressure of a topographic load is balanced only by a constant compressive strength of the crust rock, we have $h_{\rm peak} \propto g^{-1}$, and the resulting proportionality $M_{\rm cap} \propto M_p^{0.08}$ is also quite flat (though the overall basin capacity would be higher). We are being conservative about how likely planets are to have dry land by considering only dynamic topography.

It is important to emphasise that the basin capacities shown in figure \ref{fig:ocn-vol}, based on dynamic topography alone, are likely underestimating the true value. The observed topographies of Venus, Earth, and Mars produce basin capacities of 3.4, 3.3, and 2.9 Earth oceans respectively, whereas the model produces basin capacities of $\textless$1 Earth ocean. The peak and RMS elevations of our terrestrial planets are much higher than those predicted by the dynamic topography scaling here. Other mechanisms contribute to supporting higher topography on planets. Also, our model may under-predict dynamic topography for a given planet mass, as we will discuss in the next section.

\section{Discussion}
\label{sec:discussion}

\subsection{Expanding RMS topography} \label{sec:discussion-spectrum}

Figure \ref{fig:ocn-vol} suggests that reasonable changes to the spectral distribution of topography have no strong effect on how peak dynamic topography scales with planet mass, and hence on the volume of water that could be contained below this highest point. Our concern with topography's characteristic harmonic structures might thus seem somewhat tangential to (or in the worst case, distracting from) the basic problem that this study purports to address. However, these details would become more of a concern if the field can mature---and especially if we hope, someday, to use informed topography distributions as a boundary condition in exoplanet climate models \citep[e.g.,][]{turbet_habitability_2016, rushby_effect_2019}. For example, the volume calculated in (\ref{eq:ocean-integral}) represents the amount of water that would flood a planet exactly, leaving just an island with infinitesimally-small area. Yet in principle one could also calculate the maximum basin size associated with any arbitrary land fraction. These intermediate land fractions may be much more sensitive to spectral complexities, such as wide plains or anisotropic mountain ranges.

The initial questions here have justified simplified harmonic structures of topography as such. Specifically, we have presumed a log-linear model of the power spectral density, which is to say that the variance of elevation is a power-law function of the horizontal distance scale, and that this relationship is constant over the whole planet \citep[as proposed in, e.g.,][]{turcotte_fractal_1987}. Contemporary workers now know the behaviour to be much more nuanced. Local estimates of topography's spectral slope can appear notably inconstant---the surface roughness is heterogeneous---but these differences are entwined by further power laws of other statistical moments, out to virtually-infinite order, all culminating neatly in a mathematical model with three scale-invariant parameters \citep[e.g.,][]{pelletier_self-organization_1999, gagnon_multifractal_2006, lovejoy_scaling_2007, ali_saberi_percolation_2013, liucci_fractal_2017, rak_universal_2018, landais_multifractal_2019, keylock_holder-conditioned_2020}. \citet{landais_topography_2019} have demonstrated the use of such a descriptive model for synthesising surface relief of arbitrary rocky planets. Thus, the framework exists for representing full global topography layouts to a high degree of statistical realism and with few parameters. The hitch is that these parameters are empirical on a case-by-case basis: the gain in descriptive accuracy may not translate to predictive power for distant exoplanets. At present there is no theory tying the pattern to the (geophysical) process. If this gap could be bridged with more work based on Earth and solar system bodies, then these realistic mathematical models could be applied, and higher-order insight about the topographies of exoplanets might not necessarily be a fantasy.

\subsection{The role of rheology and its uncertainties} \label{sec:discussion-rheology}


Any deterministic prediction of $h_{\rm rms}$ will be hindered by the unknown mantle rheology. Increasing the activation energy of viscosity from 240~$\rm{kJ\,mol^{-1}}$ to 300~$\rm{kJ\,mol^{-1}}$ will double $h_{\rm rms}$ for an Earth-mass planet, all else being equal. This uncertainty propagation is built into our model via the scaling functional form in (\ref{eq:h_Ra_scaling}). $E_a$ enters this equation twice, in both $b$ and Ra$_i$ (via $\eta_m$). Particularly in the high-Ra$_i$ regime, small changes in the viscosity contrast parameter $b$ create large changes in $h_{\rm rms}^\prime$ (figure \ref{fig:2D-h-scale}). 

We have attempted to capture some of the rheological uncertainty by varying $E_a$ and $\eta_0$, the free parameters in the Arrhenius viscosity law (\ref{eq:eta-arrhenius}). However, we cannot claim that our results are propagating nature's true variability. Firstly, the underlying covariance of these parameters is not known. The prior range employed by our study covers only pure olivine and pure orthopyroxene, and roughly so at that. \citet{spaargaren_influence_2020} also parameterise the mineralogical control on viscosity with an extra prefactor that increases over three orders of magnitude, calibrated between ferropericlase-rich (high Mg/Si) and stishovite-rich (low Mg/Si) lower mantle compositions \citep{xu_silicon_2017, ballmer_persistence_2017}. Relating the rheological parameters to the lower or upper mantle composition in a realistic way requires not only a complex thermodynamic model predicting these mineral compositions, but also a dataset of strain rates from experiments and \textit{ab initio} mineral physics. The actual strain rate of an olivine-orthopyroxene aggregate is certainly not a simple combination of diffusion creep flow laws. Further, in practice, real mantle viscosities will be strongly sensitive to their water content, unlikely to ever be known for a given exoplanet.

The second reason why we are not capturing the true variation is that our fixed rheological model ignores structural uncertainty by design. We have only considered diffusion creep with no pressure dependence, but the creep mechanism depends on shear stress and is not known \textit{a priori}. Including pressure dependence in the parameterisation (with adiabatic profiles from an interior structure model, for example) would lead to higher viscosities and sluggish flow in the lower mantle. Importantly, and in particular for more massive planets, this fact could render the viscosity self-regulation less efficient \citep{stamenkovic_influence_2012}, meaning that internal temperatures for evolved planets become much more sensitive to initial temperature conditions, and the resulting $h_{\rm rms}$ scatters more widely (overall, retaining a hotter mantle at older ages will reduce $h_{\rm rms}$). Uncertainty would grow severer still if one allowed for complex rheological features such as a low-viscosity asthenosphere \citep{bodur_impact_2019}, which manifests in smaller-scale dynamic topography on Earth \citep{hoggard_global_2016}. Finally, technically, the lithosphere itself obeys a distinct viscoelastic rheology, and coupling these dynamics to a convection model would also modify its topography amplitudes \citep{patocka_stress_2017}---we have ignored elastic effects in this attempt (section \ref{sec:elastic}).

All this rheological uncertainty is worth discussing because dynamic topography is apparently sensitive to both viscosity's absolute value and how it changes over the boundary layers \citep{hager_long-wavelength_1989}. Low viscosities imply higher temperatures and low convective stresses. For the isoviscous case, the association of low viscosity with low topography can be seen clearly in Table 2 of \citet{lees_gravity_2020}, from which we get a numerical scaling of $h_{\rm rms}^\prime \propto \eta^{-0.6}$, with interior temperature and lithospheric thickness fixed. If we have two isoviscous layers with a stiffer top layer (i.e., approximating a cool viscous lithosphere), then there is an analytical solution for the surface normal stress induced by a spherical density anomaly at some depth \citep[equation (34) in][]{morgan_gravity_1965}. In this solution, the effect of relative viscosity is strongest when density anomalies are nearer the surface.

\subsection{Caveats to topography predictions from numerical convection}  \label{sec:discussion-modelling}

In determining a scaling relationship for the RMS and peak amplitudes of dynamic topography from numerical convection, we have assumed that details of our methodology can produce generalisable results. This section discusses some important caveats.

\subsubsection{Low-order power} 
\label{sec:discussion-loworder}
The contribution to the total power drops off quickly with spherical harmonic degree for the spectral slopes used here. Consequently, the overall RMS amplitude is unaffected by the high-frequency power, whilst the low-frequency power has a disproportionately large influence. Our simulations show a flattening-out of the topography power spectra as we go to wavelengths larger than twice the layer depth. Yet topography on Venus clearly exhibits long-wavelength features (figure \ref{fig:top-spectra}). On Earth, the dynamic topography power is largely concentrated at degree 2 \citep{hoggard_global_2016, 2021GeoJI.225.1637Y}. The relatively simple rheologies in our model cannot produce these features. Long-wavelength mantle flow on Earth may be deeply entwined with the presence of an asthenosphere and tectonic plates, themselves entwined further \citep{lenardic_toward_2019}.

Mars provides a case that's different still. Its topography is dominated by a degree-1 signal; that is, Mars shows an asymmetry where the southern hemisphere sits higher than the northern, and the volcanically-constructed Tharsis plateau dominates the east side of the former. Whilst this pattern is thought to be related to degree-1 mantle convection, as of yet there is no fully-endogenous mechanism consistent with all the observables \citep{roberts_chapter_2021}. Regardless, the processes we model will never lead to such a convection pattern. The possibility of degree-1 convection could further complicate our preliminary scaling relationship between $h_{\rm rms}$ and Ra.



\subsubsection{Geometry and heating mode effects}

Our numerical convection simulations were performed exclusively in a bottom-heated 2D box. For 2D isoviscous models, RMS topography appears consistent across Cartesian and cylindrical geometry, with a scaling exponent on Ra close to $-$1/3 as expected from theory \citep{mckenzie_convection_1974, parsons_relationship_1983}. However, in the non-isoviscous settings we study here, this scaling is not necessarily insensitive to the model geometry. It remains to be seen how higher spatial dimensions, or cylindrical or spherical geometry, would explicitly affect $h_{\rm rms}$. Internally-heated convection---best described with an altogether different formulation of the Rayleigh number---tends to result in different convective planforms and may also show different patterns with respect to dynamic topography \citep[e.g.,][]{orth_isostatic_2011}. This distinction between heating modes would be especially relevant for young planets with high radioisotope concentrations.


\subsubsection{Filtering in the lithosphere} \label{sec:elastic}

In reality, the peak amplitude of dynamic topography is modulated by the flexure of the elastic lithosphere, which depends on the lithosphere's effective elastic thickness. Thin elastic lithospheres (expected for hot stagnant lid planets such as Venus) could bring a $\lesssim 5\%$ reduction in dynamic topography \citep{golle_topography_2012, dumoulin_predicting_2013, patocka_elasticity_2019}. Here we omit this filtering for simplicity and instead predict an upper limit of dynamic topography.

In addition to these elastic effects, 
the lithosphere can deform plastically in response to convective stress, as illustrated by the crustal thickening example in figure \ref{fig:topography-schematic}b \citep{kiefer_mantle_1991,pysklywec_time-dependent_2003, zampa_evidence_2018}. We have not considered higher-order effects from the formation of a crust, whose marginally lower density with respect to mantle rock would buoy topography slightly higher.

\subsubsection{Paucity of ground truths}\label{sec:benchmarks}

Ultimately, making accurate predictions of dynamic topography amplitudes is meaningless without accurately measuring them somewhere. It is not trivial to isolate the dynamically-supported component of the cumulative topography we observe. Serious efforts at separating out the isostatic component on Venus rely on knowing the associated admittances, simulated or inferred from a crustal thickness estimate \citep{mckenzie_relationship_1994, pauer_modeling_2006, wei_gravity_2014, yang_separation_2016}, to leave a result that is not model-independent. 

For Earth, meanwhile, estimates of oceanic bathymetry less its age-depth cooling pattern can been used to navigate this impasse, revealing dynamic topography peak amplitudes of $\sim$1~km \citep{hoggard_global_2016, hoggard_oceanic_2017}. Although this result happens to align with our Earth-mass planet predictions, a direct comparison demands caution because we have been modelling stagnant lid planets---modern Earth is evidently outside this regime. Sections \ref{sec:discussion-rheology} and \ref{sec:discussion-loworder} have mentioned how the pattern of Earth's dynamic topography is a consequence of its experiencing convection under plates. Any plate behaviour is not captured in our numeric simulations. Indeed, dynamic topography observed on the only known planet with plate tectonics seems to reflect both deeper mantle flow and shallower lithospheric and aesthenospheric structure, as well as the coupling between them \citep{davies_earths_2019}. Nor is our 1D thermal history model strictly applicable: the thick, insulating lids imposed by the stagnant lid regime would lead to underestimated surface heat flow for a plate tectonics regime. Note further that this $h_{\rm peak}\sim1$~km estimate for Earth purposefully excludes the thermal bathymetry of mid-ocean ridges, a plate-scale topographic expression which could technically could fall under dynamic support.

\subsection{Caveats to using scaling relationships}

\subsubsection{Sensitivity to functional form}

A scaling law will never be more than a mathematical shortcut: a tool to preempt heavy model running for any imaginable parameter combination. This work has adopted a log-linear scaling for dynamic topography in terms of the Rayleigh number and rheological temperature scale of convection. Whilst this choice of independent parameters is indeed physically justified, it is not unique in being justifiable. 
We emphasise that the result of this study---that dynamic topography becomes essentially negligible with hotter (younger, deeper more radioactive) mantles---is fundamentally a consequence of our scaling functional form.

The interaction between $\Delta \eta$ and Ra$_i$ in our scaling somewhat complicates a comparison with previous power-law relationships for isoviscous convection---recall that constant-viscosity convection is described by a single value of the Rayleigh number. Boundary layer theory suggests that $h^\prime \sim {\rm Ra}^\gamma$ \citep{mckenzie_convection_1974, parsons_relationship_1983} with $\gamma=-1/3$, whilst more recent 3D Cartesian simulations of \citet{lees_gravity_2020} have $\gamma$ ranging from -0.289 to -0.342. Under our scaling function, an equivalent exponent to $\sim -1/3$ on Ra$_i$ is met at high values of $b \sim -23.7$, at which $h_{\rm rms}^\prime$ could be said to scale similarly to the isoviscous case. 

\subsubsection{Extrapolation across Rayleigh numbers} \label{sec:discussion-extrap}

For Ra$_1$ much greater than $3 \times 10^8$, the highest value considered in our experiments, one may be waiting prohibitively long for numerical convection models to converge. Yet the thermal histories we have produced in 1D tend to deliver these very large, out-of-range Rayleigh numbers (figure \ref{fig:1D-evolution}). Wielding the numerical scaling to estimate $h_{\rm rms}$ thus necessitates an extrapolation over up to four orders of magnitude in Ra$_i$. (Meanwhile, values of the 1D $b$ analogues are indeed accessed in 2D.) This projection into high-Ra$_i$-space has unproven fidelity, and brings a heavy caveat to our results. Namely, extrapolating scaling functions for convection rely on there being no regime change or otherwise discontinuous effects in the region to which we are blind. Yet the fitted function (figure \ref{fig:2D-h-scale}) indicates complex interactions between Ra$_i$, $b$, and $h_{\rm rms}^\prime$, which we cannot claim to have predicted in the moderate-Ra$_i$ regime, and cannot expect to predict elsewhere.

\subsubsection{Accuracy of interior Rayleigh number estimates} \label{sec:discussion-Ra}

With the above said, our Ra$_i$ results seem unrealistically high. The parameterised convection model necessitates large Ra$_i$ through its relatively hot $T_m$ and weak $\eta_m$, which viscosity self-regulation makes difficult to avoid. By comparison, mantle Rayleigh numbers used to reproduce Venus are often on the order of $\sim$10$^7$ \citep[e.g.,][]{kiefer_geoid_1992, kiefer_geoid_1998, vezolainen_timing_2003, vezolainen_uplift_2004, pauer_modeling_2006, smrekar_constraints_2012, noack_coupling_2012, huang_constraints_2013}, implying that the extrapolation issue in section \ref{sec:discussion-extrap} could in fact fix itself, if Ra$_i$ could only naturally settle down to a level a few orders of magnitude lower. However, these literature quotes come from different model setups that set Ra \textit{a priori}; e.g., to obtain desired, Earth-like average viscosities around $\sim$10$^{21}$~Pa~s. This theme of other works adopting lower Ra and higher viscosities might largely explain why our $h_{\rm rms}$ predictions appear lower \citep[e.g.,][]{kiefer_geoid_1992, huang_constraints_2013}.



Thermal models of stagnant lid planets are notorious for producing infernal mantles because their heat escape is limited by slow conduction through thick outer shells \citep[e.g.,][]{driscoll_thermal_2014}. Hence they evolve towards low viscosities and vigorous convection to aid heat loss. A parameterised model could slip into cooler temperatures by including the energetics of melting and transport of magma: likely major mantle heat sinks for stagnant lid planets \citep{moore_heat-pipe_2017, lourenco_efficient_2018}. Melting would also help to regulate mantle temperatures and viscosity because melting leads to geochemical depletion, which hinders further melting until upwelling replenishes the melt zone. Ideally, stagnant lid convection models should include melting processes. We note that melting itself also could be an important source of constructional surface topography on these planets.

\subsubsection{Model validity at high planet mass}

Rocky planets more massive than Earth can reach interior pressures high enough for perovskite to transition to post-perovskite. This phase transition, in addition to weakening the viscosity locally, could stratify the convection in the lower mantle \citep{umemoto_two-stage_2011, karato_rheological_2011, tackley_mantle_2013, umemoto_phase_2017, shahnas_penetrative_2018, ritterbex_vacancies_2018, van_den_berg_mass-dependent_2019}. Although single-layer parameterised convection models have been applied previously to massive rocky planets \citep[e.g.,][]{kite_geodynamics_2009, tosi_habitability_2017}, our model likely fails to capture the heat flow of a multi-layered system \citep{2007GeoJI.169..747V}, with potentially important implications for topography. Indeed, lower-pressure phase transitions in Earth's mantle influence its convective dynamics \citep{doi:10.1146/annurev.ea.23.050195.000433}, and including the 410-km exothermic phase change has been explicitly shown to raise dynamic topography amplitudes in convection simulations \citep{2021GeoJI.225.1637Y}.

\subsection{A crustal strength limit and the inundation of the TRAPPIST-1 system}

\begin{figure*}
    \centering
    \plotone{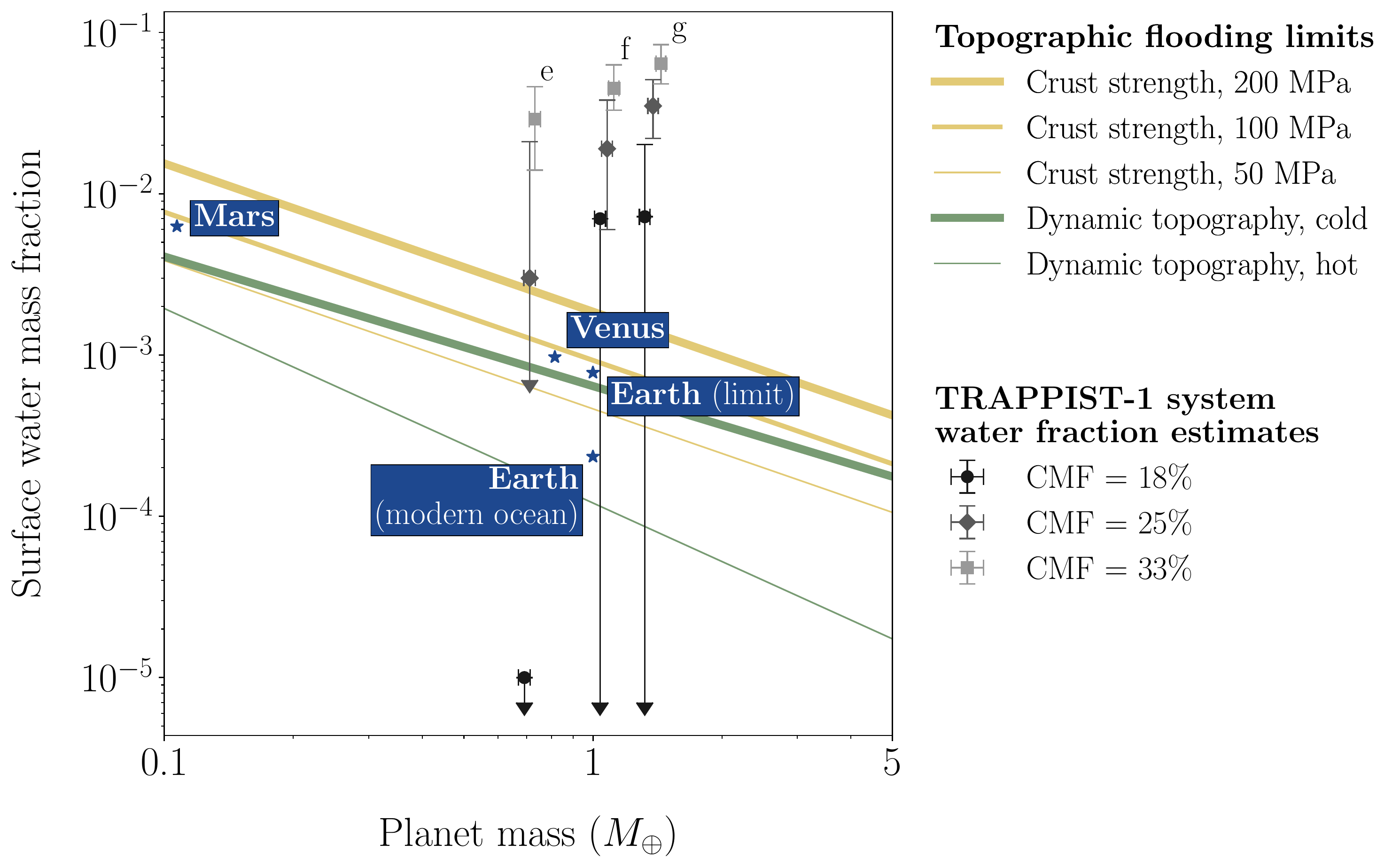}
    \caption{Various scalings for the maximum surface water capacity set by a planet's peak elevation, expressed as a fraction of the total planet mass. The yellow lines show the peak topography balanced by crustal rock strength alone, and scales approximately with $M_p^{-0.9}$; line widths correspond to different assumptions about the maximum strength with a fixed crust density of 2700~$\rm{kg\,m^{-3}}$. The thick green line shows pure dynamic topography with the coolest mantles considered, given a dry olivine rheology ($\propto M_p^{-0.8}$). The thin green line is the same for the hottest mantles ($\propto M_p^{-1.2}$). Scalings assume the mass-radius relation in (\ref{eq:MR}) and a red noise-like topographic spectral structure. Points with error bars are estimates of the surface water inventories of planets e--g in the TRAPPIST-1 system from \citet{agol_refining_2021}, for different possible values of the core mass fraction (CMF). Note that their analysis suggests cores most likely smaller than the Earth-like CMF of 33\%. Our thermal evolution model does not include tidal heating, which would push the TRAPPIST-1 planets towards higher mantle temperatures. For context, the labelled blue stars show the maximum ocean masses that could be contained on Venus, Earth and Mars, plus Earth's actual ocean mass.}
    \label{fig:max-ocn}
\end{figure*}

\citet{agol_refining_2021} give preliminary constraints on the surface water content of the TRAPPIST-1 planetary system, for different values of the CMF and assuming all water exists as a condensed surface layer. Although the problem is degenerate, planets e--g appear consistent with water layers deeper than Earth's, on the order of at least 0.1\% of the planet mass. Other independent estimates have produced similar results \citep{acuna_characterisation_2021}.  
This water budget would place TRAPPIST-1e to g well above the upper water mass limit for maintaining land with dynamic topography. Note, however, that the high rates of tidal heating inferred for some of these planets \citep{barr_interior_2018} would reduce dynamic topography beyond what is modelled here.

As we have previously emphasised, however, the true limit to elevation differences on a planet will be higher than that suggested by purely dynamic topography. To estimate a planet's total scope for land, we can calculate the minimum value of $h_{\rm peak}$ required for an instance of land on a planet with a given radius and surface water content. We find that any instance of land on TRAPPIST-1e would require a peak topographic amplitude of $\sim$40~km (a minimum RMS topography of $\sim$10~km), given 0.3 wt.\% surface water (\citeauthor{agol_refining_2021}'s estimate for a CMF of 0.25). Then one could compare this minimum to a rough estimate of the overall maximum elevation. 

In section 1 we motivated a crustal strength limit: for a surface load of $\rho g h$, somewhere in the crust below, at a depth of about 1/4 times the load width, a minimum stress difference $Y$ of 1/2 to 1/3 $\rho g h$ is sustained \citep{jeffreys_earth_1929}. This result assumes a flat earth model of elastic stress distributions, and holds for various load configurations of horizontal scale less than a few hundred kilometres. \citet{melosh_planetary_2011} illustrates that the force balance given by
\begin{equation}\label{eq:crust_strength}
    Y \approx 0.5 \rho_c g h,
\end{equation}
with a crust density $\rho_c = 2700$~$\rm{kg\,m^{-3}}$, and $Y$ set at an effective crustal strength on the order of 100 MPa, will roughly reproduce the maximum elevations of Venus, Earth, and Mars (figure \ref{fig:max-ocn}). Whilst this estimate is certainly an oversimplification, a more rigorous effort will naturally become very complicated, not the least due to the difficulty in predicting, from planetary bulk properties, a value of $Y$ corresponding to the maximum $h$.

In typical crustal strength models, the strength increases with depth (lithostatic pressure) according to the rock's resistance to frictional sliding in the relatively cool, shallow part---the brittle regime---until viscosity is low enough to favour ductile deformation instead, and strength starts to decrease with depth (temperature). Thus the strongest part of the crust is near this brittle-ductile transition. However, the resulting strength maxima of $\sim$500~MPa or more for Earth-like conditions \citep{katayama_strength_2021} would imply $\sim$40 km of peak topography using (\ref{eq:crust_strength}); it is a limit not necessarily reached in practice. Further complicating the application of (\ref{eq:crust_strength}), crustal strength profiles are strongly sensitive to the temperature profile and porosity of the crust---generating these profiles for arbitrary exoplanets must attend to assumptions on these facets \citep{byrne_effects_2021}---and surface gravity has a nonlinear effect on brittle strength through its influence on porosity and fracture density \citep{heap_low_2017}. For example, doubling the thermal gradient will approximately halve the maximum $Y$---and thus $h$---in a dry case, and including hydrostatic pore fluid pressure shows a similar decrease \citep{katayama_strength_2021}.

A parallel approach to estimating maximum elevation differences from crustal concerns comes from isostasy. The height of a topographic feature above a plain is $h_A = (t_R - t_{\rm avg})(\rho_m - \rho_c)/\rho_c$, where $t_R$ is the thickness of the crust below the feature and $t_{\rm avg}$ is the average crustal thickness of the plain. For a basaltic crust (the primary crust formed from an Earth-like bulk composition), the maximum value of $t_R$ is set by the phase transition from basalt to denser and unstable eclogite: the crust cannot be much thicker than the depth of this transition. This fact limits the peak isostatic $h_A$ to about 15 km for a Venus-like case \citep{jull_implications_1995}. However, the depth of this phase transition depends sensitively on the crust thermal structure, and estimating $h_A$ in practice requires knowing $t_{\rm avg}$.

Finally, the height limits of volcanoes in particular must follow tighter rules. Magma will only rise to the top of a vent---and contribute to a growing pile of lava---so long as the vertical pressure gradient across the system is positive. \citet{castruccio_influence_2017} write this limit as $h_{\rm max} = (\Delta\rho/\rho_m) H + \Delta P_i/(\rho_m g)$, where $\Delta\rho$ is the density contrast between the crust and the magma, $H$ is the depth from the surface to the magma chamber, and $\Delta P_i$ is the critical overpressure to trigger an eruption (the pressure that would crack the magma chamber roof, related to the tensile strength of the crust and tellurically on the order of $\sim$20 MPa). Although a narrow range of $H$ can be argued for on Earth, related to the magma water content and crustal rheology \citep{huber_optimal_2019}, this concept has not yet been expanded to comparative planetology. 

In light of the above complexities, it is difficult to find a middle ground between the oversimplification of (\ref{eq:crust_strength}) using a universal crustal strength estimate, and a careful case-by-case application. 
We will employ the former for the present purpose of comparing peak dynamic topography to peak total topography. We consider $Y = 100$ MPa, ostensibly representing the compressive strength of granite---the difference in compressive strength between an average granite and average basalt seems to be smaller than the spread seen across individual basalt samples in various laboratory conditions \citep{heap_low_2017}---but also include scalings for half and double this strength value.

Figure \ref{fig:max-ocn} plots the containable ocean mass fraction scalings corresponding to both this crust strength limit and to the dynamic topography limits calculated previously. For a given scaling relationship, points above the line would be waterworlds. We see that planet e may have coexisting land and water if its crust could withstand around 200 MPa of normal stress. Although these strengths can be achieved on Earth, it is not immediately obvious that they would available at the right loci.
Note that it is very difficult in practice to put a lower limit on these water budgets. Nevertheless, according to \citet{agol_refining_2021}, TRAPPIST-1e through g could easily be wet enough that estimating their land propensities may seem moot. However, our growing catalogue of planets may soon present a case study closer to the waterworld-land world transition. 

Another takeaway from figure \ref{fig:max-ocn} is that for the most massive rocky planets, amplitudes of dynamic topography in the most favourable case seem to approach the overall limit. Scalings for different internal heating scenarios have different slopes because, as surface heat fluxes increase with the surface area-to-volume ratio, larger planets are penalised such that any extra radiogenic heat would escape less easily. Thus more internal heating per unit volume in more massive planets will have a more drastic effect on topography.

At the moment, it is not guaranteed that constraints on these or any rocky exoplanet water budgets could be tightened much in the future. With current Bayesian inference methods, uncertainties on retrieved water mass fractions may be capped at around $\sigma \approx 10$~wt.\%, independent of the observational uncertainty on the planet mass itself \citep{otegi_impact_2020}. Meanwhile, topography can avert waterworlds only for water mass fractions of $\lesssim 1$~wt.\%. Therefore, with respect to predictions about a given exoplanet, any topographic contribution to land coverage could be washed out by the uncertainty on the inferred water budget.

\subsection{Constraints from astrophysical data} \label{sec:obvs}

In figure \ref{fig:1D-h-scale}, we predicted how dynamic topography might vary as a function of several properties broadly deemed observable. None of these properties will be perfectly known, or even necessarily constrained well-enough such that they are not the dominant source of uncertainty, but we will leave a more detailed assessment of this uncertainty to future work.

In any case, an obvious fact emerging from our scaling law application is that there is a pivotal future role to be filled for any constraints on rocky planet compositions. This study provides yet another example of how higher-order properties of planetary interiors govern their surface character. Namely, mantle viscosities, radiogenic heating rates, and core mass fractions all relate to planetary ratios of certain major elements: viscosities decrease with Mg/Si, radiogenic heating rates increase with U/Si and Th/Si, and core mass fractions increase with Fe/O. Exoplanet compositional parameters are not completely inaccessible because refractory element ratios are expected to generally preserve themselves between a star and its planets \citep{thiabaud_elemental_2015, hinkel_starplanet_2018, putirka_composition_2019, adibekyan_chemical_2021}. Although pilot work is surely needed, this useful fact means that element abundances from stellar spectra offer a promising constraint on planetary interior dynamics. Additionally, measurements of the same element ratios in polluted white dwarf spectra could inform the underlying natural distributions of bulk rocky planet composition across nearby star systems \citep{bonsor_host-star_2021}.

Observables for exoplanetary topography itself would be buried quite deep. \citet{mctier_finding_2018} proposed that extreme topographic features could induce scatter in an exoplanet's transit photometry, but the associated signal would not be detectable with realistic photometric precision. Proposed next-generation direct imaging missions might be capable of enough precision for the exo-cartography of small planets---solving the inverse problem of 2D albedo distributions from time-resolved light-curves---which might discriminate between land and ocean surfaces \citep{cowan_mapping_2018, farr_exocartographer_2018, lustig-yaeger_detecting_2018, kawahara_global_2020, aizawa_global_2020}. Interpretations of the data may remain highly model-dependent and burdened by cloud removal, however \citep{paradise_fundamental_2021, teinturier_mapping_2022}. Ocean fractions might also be discerned from near-infrared polarimetric observations \citep{takahashi_polarimetric_2021}. A land fraction between zero and unity would necessitate some surface roughness, leading to an upper limit on the water budget given some inferences about topographic propensity. 



\section{Conclusions}

This work has predicted scaling relationships for the RMS amplitude of dynamically-supported topography on stagnant lid planets, which we propose to be a deterministically-tractable aspect of rocky exoplanet surface character. We find RMS topography to decrease strongly with higher interior temperatures and lower mantle viscosities. Planets near the upper mass-limit of rockiness would thus have inconsequential dynamic topography, as would planets with radioisotope abundances several times that of Earth. For planets less than about twice the mass of Earth, our thermal history model predicts RMS dynamic topography on the order of hundreds of metres. This result emphasises that modelling purely dynamic topography will underestimate a planet's true RMS elevation. A robust upper limit to total topography may be limited by our ability to predict crustal thicknesses.

Considering that dynamic topography is guaranteed to exist on active planets, however, the model can be used to infer, with strong caveats, whether subaerial land exists on a planet for a given surface water budget. We define the ocean basin capacity as the volume of water that could be contained below the highest elevation. As planet size increases, interior temperatures and surface gravity increase and topography shrinks, but the available storage of the ocean basins expands with the surface area. These effects nearly cancel out at Earth-like radiogenic heating rates, leading to a constant ocean basin capacity of about 0.3 Earth oceans if topography is dynamically-supported alone. For a 1-$M_\oplus$ planet this translates to a maximum surface water mass fraction of $\sim$60~ppm before the planet has no land above sea level. The same water budget would flood more massive planets. In reality, volcanic construction would lead to higher surface relief than that from dynamic topography alone---in modelling only the latter, we are providing a lower limit, or ``worst-case scenario," of the true ocean basin capacity. To avert waterworlds on high mass planets, other sources of topography would be vital.




A useful waypoint from this work is a naive scaling relationship of RMS dynamic topography in terms of the mantle Rayleigh number and viscosity contrast, for chaotic time-dependent convection with large viscosity contrasts. Our results suggest a weaker Ra-dependence and overall higher topography amplitudes compared to the isoviscous convection scalings previously reported. 

Segments of the general approach here might guide other mysteries about rocky planet surface architecture---which seems, at the time of writing, an unpopulated but fertile field of research. We conceive of a framework into which new geophysical or geomorphological models could easily slide. Particularly, the method of gauging whole surface layouts via the RMS amplitude extends to other ways of generating large-scale topography, so long as---and this step is nontrivial---one could write process-based scaling laws for how its RMS value changes with planetary bulk properties. Reasonable assumptions about the power spectral distribution of topography give peak amplitudes between 3.5 and 3.9 times the RMS value, consistent across different ways of supporting loads. With that said, care should be taken to not overemphasise the general feasibility of such applications, given that decades of examination into our own planet's topography have not yet reached any steadfast deterministic rules. To push the marriage between these sciences further \citep{shorttle_why_2021}, then, finding tighter links between pattern and process on the surface of Earth will be paramount to understanding how landscapes manifest on billions of rocky planets in the universe.

\vspace{2cm}
We acknowledge the support of the University of Cambridge Harding Distinguished Postgraduate Scholars Programme and the Natural Sciences and Engineering Research Council of Canada (NSERC). Cette recherche a \'{e}t\'{e} financ\'{e}e par le Conseil de recherches en sciences naturelles et en g\'{e}nie du Canada (CRSNG). We thank the Computational Infrastructure for Geodynamics (geodynamics.org) which is funded by the National Science Foundation under award EAR-0949446 and EAR-1550901 for supporting the development of ASPECT.

\newpage

\appendix
\restartappendixnumbering

\section{Spherical harmonic methods for topography}
\label{sec:sph-harms}

\subsection{A baseline power spectrum} \label{sec:spectral-model}

We choose our Case 4 simulation (Table \ref{tab:aspect}) from which to extract a scaleable model spectrum of the surface dynamic topography, since its temporal distribution of $h_{\rm rms}^\prime$ is the most narrow. A type-2 orthonormalised discrete cosine transform of this profile produces a Fourier representation,
\begin{align}
    \begin{split}
    f_p &= 2 \gamma \sum_{n=0}^{N-1} h_n^\prime \cos \left( \frac{\pi p (2n + 1)}{2N}\right),\\
    \gamma &= 
    \begin{cases}
      \sqrt{\frac{1}{4N}}, & \text{if}\ p=0 \\
      \sqrt{\frac{1}{2N}}, & \text{otherwise,}
    \end{cases}
\end{split}
\end{align}
from which we can find a 1D power spectral density,
\begin{equation}
    \phi^{\rm 1D}_0 = 2 \Delta x^\prime \left(f_p\right)^2,
\end{equation}
as a function of dimensionless wavenumber,
\begin{equation}
    k^\prime = \frac{\pi}{L^\prime} p,
\end{equation}
where $h_n^\prime$ is the height of dynamic topography at sample point $n$, $N$ is the number of sample points in the spatial profile (fixed by the mesh size), $p = [0, ..., N - 1]$, $L^\prime = 8$ is the dimensionless box width, and $\Delta x^\prime = L^\prime / N$. We calculate $\phi^{\rm 1D}_0$ at every model time step and use the average for our baseline spectrum. This spectrum has an RMS amplitude $h^{\prime}_{\rm rms, 0}$. 

There is an upper wavenumber limit, $k^\prime_{\rm max}$, at around the equivalent wavelength of the upper thermal boundary layer thickness, $\delta_{\rm rh}$, where features narrower than this are not meaningful for the dynamic topography. We also observe all spectra roughly rolling off to a constant value at wavenumbers below around twice the convection cell depth, so we set $k_{\rm min}^\prime = 2d$. In log-log space, $\phi^{\rm 1D}_0$ is approximately linear from $k_{\rm min}^\prime$ to $k^\prime_{\rm max}$. Therefore we approximate the power spectra by two line segments. We fit a constant slope between $k^\prime_{\rm min}$ and $k^\prime_{\rm max}$, and assign a value of $\phi^{1D}_0(k^\prime_{\rm min})$ wherever $k^\prime < k^\prime_{\rm min}$. This fit is done to the average power spectral density over all time steps for the given simulation. We interpolate this fitted function such that it has a discrete value at each integer spherical harmonic degree $l$, where $l = k^\prime R_p^\prime - 0.5$, from $l=1$ to the nearest degree to $k^\prime_{\rm max}$. That is, we do not scale $k^\prime_{\rm max}$. Whilst realistically $k^\prime_{\rm max}$ would increase with Ra$_1$, the effect on $h_{\rm rms}^\prime$ is small (less than one part in a thousand) because these high wavenumber bands hold such little relative power. For this generic spectrum we assume a dimensionless planet radius $R_p^\prime = 2$ (a core radius fraction of 0.5 for a dimensionless mantle depth of 1; varying $R_p^\prime$ has negligible effects on the results).

\begin{figure}
    \centering
    \epsscale{0.7}
    \plotone{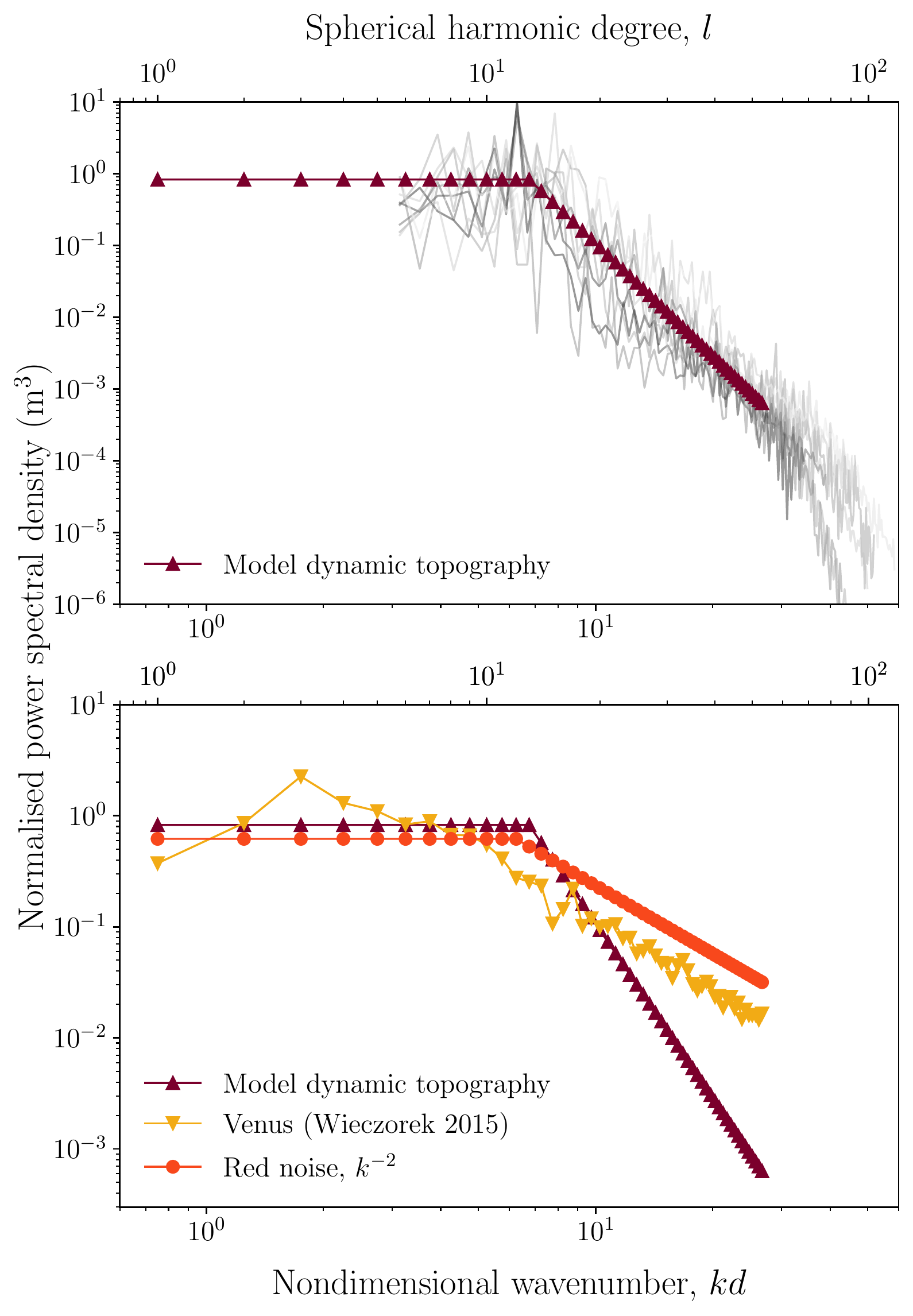}
    \caption{\textit{(Top:)} Dimensionless 1D power spectral densities of dynamic topography from 2D numerical convection simulations, normalised to an RMS power of unity. In purple triangles is the model dynamic topography spectrum obtained from a log-linear fit to the Ra$_1 = 10^8, \Delta \eta = 10^7$ case. \textit{(Bottom:)} The model dynamic topography spectrum shown with, in yellow triangles, the observed overall topography of Venus \citep{wieczorek_gravity_2015}, and, in red circles, a theoretical spectrum with a power-law dependence $\propto k^{-2}$, corresponding to red noise.
    \label{fig:top-spectra}}
\end{figure}

Figure \ref{fig:top-spectra} shows the 1D power spectral densities $\phi^{\rm PSD}_h$ of dynamic topography computed from our 2D numerical modelling experiments, normalised as a percentage of the total power. Between $k_{\rm min}^\prime$ and $k_{\rm max}^\prime$, the log-linear slopes of the topography spectra are roughly similar within the noise for all Ra$_1$, $\Delta \eta$ cases. Due to our limited number of 2D runs, however, we cannot really make a compelling case for this statement, and we would not back our interim result outside of its intended, rather inconsequential usage here. For example, we might expect more vigorous, higher-Ra convection to exhibit more smaller-scale drips from the upper thermal boundary layer, leading to slightly more topographic power at high wavenumbers---although the total power would be virtually unaffected by these high-frequency features. Note also that because the spatial domain topography is 1D, data paucity will always entail a certain amount of noise, compared to a 2D grid of topography from a 3D convection simulation.

Also in figure \ref{fig:top-spectra} is the observed topography spectrum of Venus from \citet{wieczorek_gravity_2015}. On Venus, elastic and compositional sources of topography are superimposed upon dynamic topography. Venus' spectrum thus provides an empirical modification of the pure dynamic topography. As a third and final spectral model, we have the theoretical red noise spectrum given by the power law $\phi^{\rm PSD}_h \propto k^{-2}$ and a roll-off wavenumber the same as the numerical spectrum. Compared to the numerical dynamic topography, Venusian topography and red noise topography both have a shallower slope and retain more power at higher wavenumbers---as expected from the high-frequency nature of topography created by impact cratering and volcanism. The Venus spectrum additionally shows a peak at degree $l=3$. Note that these (normalised) spectra represent different geophysical and geomorphologic processes, and are therefore not expected to have the same absolute RMS value.

\subsection{Generating random maps} \label{sec:spectral-repeat-top}

\begin{figure}
    \centering
    \epsscale{0.6}
    \plotone{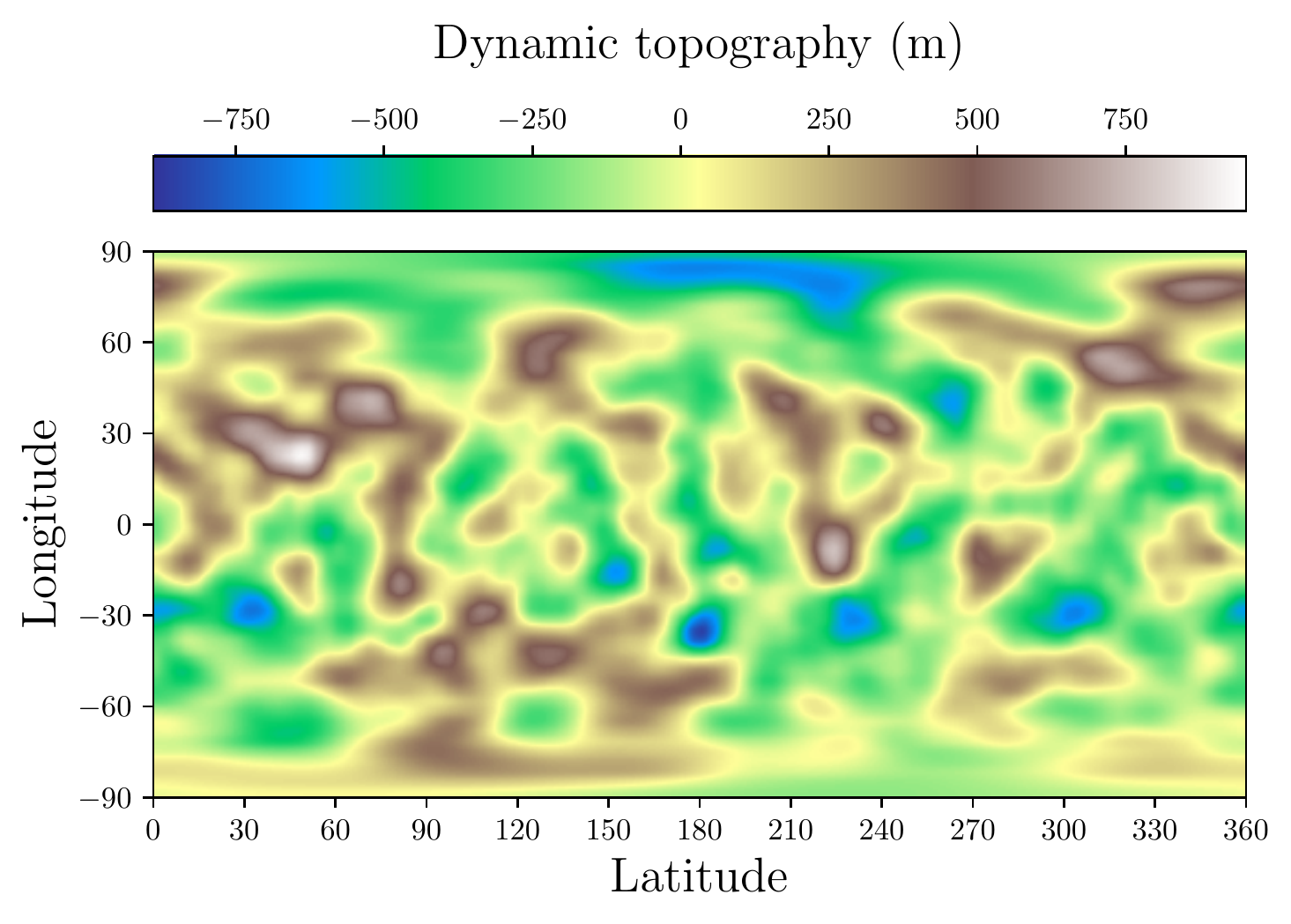}
    \caption{A synthetic topography map, obtained from a random power spectrum ($l_{\rm max} = 53$) consistent with the numerically-modelled ``baseline" dynamic topography spectrum (see text for details on randomisation). This map has a peak elevation of 820 m and an RMS elevation of 190 m. The nominal planet has a mass of 1~$M_\oplus$, dry olivine rheology, and a solar radiogenic heating budget.
    \label{fig:synth-map}}
\end{figure}

We use the {\tt pyshtools.SHCoeffs.from{\_}random()} function to obtain a set of spherical harmonic coefficients consistent with $\phi^{1D}_0$ \citep{wieczorek_shtools_2018}. This function requires a power per $l$ (dimensional units m$^2$), so we apply a conversion from $\phi^{1D}_0$ (dimensional units $\rm{m^2\,m}$). First we find the effective 2D power spectral density assuming radial symmetry, $\phi^{2D}_{\rm iso}$ (dimensional units $\rm{m^2\,m^2}$), which would correspond to our 1D spectrum:
\begin{equation}
\phi^{2D}_{\rm iso} = \frac{1}{k^\prime} \phi^{1D}_0.
\end{equation}
The power per $l$ is:
\begin{equation}
S_l = \frac{\phi^{2D}_{\rm iso} \left(2 l + 1\right)}{4 \pi R_p^{\prime 2}}.
\end{equation}
With these normalisations, the total power per coefficient,
\begin{equation}
S_{lm} = \frac{S_l}{2l + 1},
\end{equation}
is proportional to $\phi^{2D}_{\rm iso}$. In converting our spectra into 2D equivalents, we are presupposing that 2D Cartesian and 3D spherical models result in approximately similar topography power spectra with consistent $h_{\rm rms}^\prime$. Using the output from \citet{lees_gravity_2020}, we have verified that constant-viscosity convection in Cartesian geometry indeed produces similar spectra between 2D and 3D, but the assumption remains a caveat until dedicated 3D spherical realisations can test it. Nevertheless, we already know that it is incorrect to try fitting a scaling function to 2D numerical $h_{\rm peak}$ directly---this quantity is certainly sensitive to details of the model setup, as we have mentioned in section \ref{sec:methods-hscaling}.

If we are seeking a spatial map of a hypothetical spectrum other than $\phi^{1D}_0$ (i.e., different RMS value), we take advantage of the fact that numerical dynamic topography spectra will appear to have roughly consistent slopes between $k^\prime_{\rm min}$ and $k^\prime_{\rm max}$, and hence scale $S_l$ appropriately, 
\begin{equation}
\bar{S_l} = S_l \left(\frac{h^{\prime}_{{\rm rms}, 1}}{h^\prime_{{\rm rms}, 0}}\right)^2,
\end{equation}
where $h^{\prime}_{{\rm rms}, 1}$ refers to the desired rms of the new spectrum. 

We can now obtain our set of coefficients via {\tt pyshtools}: 
random spherical harmonic coefficients are generated from a normal distribution with unit variance, subject to the strong assumption of no correlation between wavenumbers. 



Then we again use {\tt pyshtools} to expand the random spherical harmonic coefficients onto a Gauss-Legendre quadrature grid.
At this stage we can dimensionalise the spatial domain topography with (\ref{eq:dimensionalise}), given the results of the parameterised convection model. A sample elevation map is shown in figure \ref{fig:synth-map}. 
Because the randomly-generated spherical harmonic coefficients are not unique for a given power spectrum, we reduce the noise by generating 500 sets of coefficients and taking the average of the resulting peak elevation values.

\section{Tabular output of 2D numerical convection experiments}
\label{sec:aspect-table}

Table \ref{tab:aspect-out} provides additional numerical output. See section \ref{sec:methods-numerical} for definitions of these quantities. Nu is the Nusselt number, the ratio of convective to conductive heat transfer at the surface, calculated as ${\rm Nu} = Y^\prime F^\prime_0 / [k^\prime(T^\prime_1 - T^\prime_0)]$, where $Y^\prime$ is the dimensionless box height, $ F^\prime_0$ is the total surface dimensionless heat flux divided by the dimensionless box width, $k^\prime = 1$ is the dimensionless thermal conductivity, and $T^\prime_1$ and $T^\prime_0$ are the dimensionless temperatures at the bottom and top boundaries respectively.

\begin{table}[h!]
\centering
\caption{Selected time-averaged results of the numerical model. Symbols are defined in the text. \label{tab:aspect-out}}
\footnotesize
\begin{tabular}{@{} c r r r r r r r r r r r @{}}
\toprule
Case & Ra$_1$ & $\Delta \eta$ & Ra$_i$ & $\delta_{\rm lid}^\prime$ & $\delta_{\rm rh}^\prime$ & $T_i^\prime$ & $T_{\rm lid}^\prime$ & $\Delta T_{\rm rh}^\prime$ & Nu & $h_{\rm rms}^\prime$ & $h_{\rm peak}^\prime$  \\
\midrule

2 & $2 \times 10^{8}$ & $1 \times 10^{6}$ & $7.20 \times 10^{7}$ & $0.133$ & $0.0248$ & $0.926$ & $0.785$ & $0.141$ & $6.17$ & $0.00716$ & $0.0152$ \\
3 & $3 \times 10^{8}$ & $1 \times 10^{6}$ & $1.07 \times 10^{8}$ & $0.118$ & $0.0218$ & $0.925$ & $0.790$ & $0.135$ & $6.97$ & $0.00667$ & $0.0130$ \\
4 & $1 \times 10^{8}$ & $1 \times 10^{7}$ & $3.62 \times 10^{7}$ & $0.199$ & $0.0370$ & $0.937$ & $0.794$ & $0.143$ & $4.10$ & $0.00893$ & $0.0214$ \\
5 & $2 \times 10^{8}$ & $1 \times 10^{7}$ & $7.08 \times 10^{7}$ & $0.165$ & $0.0238$ & $0.936$ & $0.816$ & $0.120$ & $5.12$ & $0.00610$ & $0.0159$ \\
6 & $3 \times 10^{8}$ & $1 \times 10^{7}$ & $1.07 \times 10^{8}$ & $0.148$ & $0.0215$ & $0.936$ & $0.816$ & $0.120$ & $5.70$ & $0.00673$ & $0.0145$ \\
7 & $1 \times 10^{8}$ & $1 \times 10^{8}$ & $3.60 \times 10^{7}$ & $0.235$ & $0.0394$ & $0.945$ & $0.806$ & $0.138$ & $3.50$ & $0.00907$ & $0.0243$ \\
8 & $2 \times 10^{8}$ & $1 \times 10^{8}$ & $7.24 \times 10^{7}$ & $0.199$ & $0.0295$ & $0.945$ & $0.821$ & $0.124$ & $4.23$ & $0.00765$ & $0.0174$ \\
9 & $3 \times 10^{8}$ & $1 \times 10^{8}$ & $1.08 \times 10^{8}$ & $0.179$ & $0.0253$ & $0.945$ & $0.826$ & $0.118$ & $4.75$ & $0.00788$ & $0.0179$ \\
10 & $1 \times 10^{8}$ & $1 \times 10^{9}$ & $3.57 \times 10^{7}$ & $0.274$ & $0.0427$ & $0.950$ & $0.819$ & $0.131$ & $3.03$ & $0.00815$ & $0.0252$ \\
11 & $2 \times 10^{8}$ & $1 \times 10^{9}$ & $7.20 \times 10^{7}$ & $0.232$ & $0.0329$ & $0.951$ & $0.831$ & $0.120$ & $3.65$ & $0.00878$ & $0.0250$ \\
12 & $3 \times 10^{8}$ & $1 \times 10^{9}$ & $1.11 \times 10^{8}$ & $0.213$ & $0.0262$ & $0.952$ & $0.846$ & $0.105$ & $4.07$ & $0.00876$ & $0.0180$ \\


\bottomrule
\end{tabular}
\end{table}

\bibliographystyle{aasjournal}
\bibliography{main.bib}
\end{document}